\documentclass[useAMS,usenatbib,usegraphicx]{mn2e}
\usepackage{graphicx}
\usepackage[active]{srcltx}
\usepackage{times}
\newcommand{\sauron}{{\texttt {SAURON}}}
\newcommand{\Xsauron}{{\small {XSAURON}}}

\def\aj{AJ}             	
\def\araa{ARA\&A}       	
\def\apj{ApJ}           	
\def\apjl{ApJ}          	
\def\apjs{ApJS}         	
\def\aap{A\&A}          	
\def\aaps{A\&AS}        	
\def\mnras{MNRAS}       	
\def\pasp{PASP}         	
\def\pasj{PASJ}         	

\newcommand{\hb}{H$\beta$}

\newcommand{\mgb}{Mg\,$b$}

\newcommand{\fe}{Fe 5015}

\newcommand{\oiii}{[O{\small III}]}

\def\oiii{[O\,{\sc iii}]}

\def\hbeta{\hbox{H$\beta$}}

\def\hii{\hbox{H\,{\sc ii}}}

\title[The \sauron\ project.~XI] {The SAURON project - XI.
Stellar Populations from Absorption Line Strength Maps of 24 Early-Type Spirals}

\author[Peletier et al.] 
{Reynier F. Peletier$^{1}$\thanks{peletier@astro.rug.nl}, Jes\'us
Falc\'on-Barroso$^{2,3}$, Roland Bacon$^{4}$, Michele Cappellari$^{5,3}$,  
\newauthor 
Roger L. Davies$^{5}$, P.T. de Zeeuw$^{3}$, Eric Emsellem$^{4}$, 
Katia Ganda$^{1}$, Davor Krajnovi\'c$^{5}$, 
\newauthor
Harald Kuntschner$^{6}$, Richard M. McDermid$^{3}$, 
Marc Sarzi$^{7}$, Glenn van de Ven$^{8,3}$\\
$^1$Kapteyn Astronomical Institute, University of Groningen, P.O. Box 800, 9700 AV Groningen,
The Netherlands\\
$^2$European Space Agency / ESTEC, Keplerlaan 1, 2200 AG Noordwijk, The Netherlands\\
$^3$Sterrewacht Leiden, University of Leiden, Niels Bohrweg~2, 2333~CA Leiden, The Netherlands\\
$^4$Universit\'e de Lyon 1, CRAL, Observatoire de Lyon, 9 av. Charles Andr\'e,
F-69230 Saint-Genis Laval; CNRS, UMR 5574 ; ENS de Lyon, France\\
$^5$Sub-Department of Astrophysics, University of Oxford, Denys Wilkinson
Building, Keble Road, Oxford OX1 3RH, United Kingdom\\
$^6$Space Telescope European Coordinating Facility, European Southern
Observatory, Karl-Schwarzschild-Str~2, 85748 Garching, Germany\\
$^7$Centre for Astrophysics Research, University of Hertfordshire, Hatfield, UK\\
$^8$Institute for Advanced Study, Einstein Drive, Princeton, NJ 08540, USA
}

\begin{document}
\maketitle
%
%
\begin{abstract}
We present absorption line strength maps of a sample of $24$ representative
early-type spiral galaxies, mostly of type Sa, obtained as part of the {\tt
SAURON} survey of nearby galaxies using our custom-built integral-field
spectrograph.  Using high-quality spectra, spatially binned to a
constant signal-to-noise, we measure several key age, metallicity and abundance
ratio sensitive indices from the Lick/IDS system over a contiguous
two-dimensional field including bulge and inner disc.  We present maps of \hb,
\fe, and \mgb, for each galaxy. We find that Sa galaxies on the average
have slightly smaller \mgb\ and \fe\ line strengths than ellipticals and S0s,
and higher \hb\ values, but with a  much larger scatter. 

The absorption line maps show that many galaxies contain some younger
populations ($\leq$~1~Gyr), distributed in small or large inner discs, or in
circumnuclear star forming rings.  
In many cases these young stars are formed in circumnuclear mini-starbursts,
which are dominating the light in the centres of some of the early-type spirals. These
mini-starburst cause a considerable scatter in index-index diagrams such as
\mgb\ -- \hb\ and \mgb\ -- \fe, more than is measured for early-type galaxies.
We find that the central regions of Sa galaxies display a wide range in ages, 
even within the galaxies. We find that the central regions of 
early-type spirals are often dusty, with a good correlation between the
presence of young central stellar  populations and a significant amount of dust
extinction. 50\% of the sample show velocity dispersion drops in their centres.

All of the galaxies of our sample lie on or below the \mgb\ -- $\sigma$
relation for elliptical galaxies in the Coma cluster, and above the \hb\
absorption line -- $\sigma$ relation for elliptical galaxies. If those
relations are considered to be relations for the oldest local galaxies we see that our
sample of spirals has a considerable scatter in age, with the largest scatter
at the lowest $\sigma$. This is in disagreement with highly inclined samples, in which
generally only old stellar populations are found in the central regions.

The discrepancy between our sample and highly inclined samples, and the presence of
so many stellar velocity dispersion dips, i.e., so-called $\sigma$-drops, in these spiral galaxies with large bulges (type Sa) can be understood if the
central regions of Sa galaxies contain at least 2 components:  a thin, disc-like
component, often containing recent star formation, and another, elliptical-like 
component, consisting of old stars and rotating more slowly, dominating the light
above the plane.
These components together form the photometrically defined
bulge,  in the same way as the thin and the thick disc co-exist in the solar
neighbourhood. In this picture, consistent with the  current literature, part of the
bulge,  the thicker component, formed a very long time ago. Later, stars continued to
form in the central regions of the disc, rejuvenating in this way the bulge through
dynamical processes.  This picture is able to explain in a natural way the
heterogeneous stellar populations and star formation characteristics that we are
seeing in detailed observations of early-type spiral galaxies.

\end{abstract}
%
%

\section{INTRODUCTION}
\label{sec:intro}

The measurement of absorption line strengths in combination with stellar population models
has been used for many years to probe the age, metallicity and abundance ratios of certain
elements in integrated  stellar populations of galaxies (e.g., Faber 1973, Burstein et al.
1984, Gonz\'alez 1993, Davies et al. 1993, Kauffmann et al. 2003). Up to recently, this was
done using long-slit spectroscopy. Now  two-dimensional absorption line maps are starting to
emerge. Kuntschner et al. (2006, Paper VI)  recently published line strength maps for the {\tt SAURON}
sample of 48 ellipticals  and S0s. One of the main advantages of having maps is that one
can easily  identify two-dimensional structures, so that one can isolate areas of e.g., 
high Mg/Fe ratio, or young stellar populations, and then understand better their origin by
looking at their morphology or kinematics.  Kuntschner et al. used the Lick/IDS system (Faber et
al. 1985, Gorgas et al. 1993,  Worthey et al. 1994), in order to allow an easy comparison
with existing data.  Here we present similar data, but for the inner regions of the {\tt SAURON}
sample of Sa galaxies. Since this type of spiral galaxies consists, amongst others, of a bulge 
and an exponential disc, one might wonder which fraction of the bulge is covered by the {\tt
SAURON} field. In this paper we will use the practical definition, which has been often used in
the literature, that the bulge is the central component causing the excess in light above the
exponential disc (e.g. MacArthur et al. 2003, de Jong 1996, Graham 2002).
This so-called bulge is often much brighter than the disc in the centre of the galaxy (up to
factors of 100 or more). In a following paper (Falc\'on-Barroso et al., in preparation) we will present a photometric
bulge-disc decomposition for this sample, into a S\'ersic bulge and an exponential disc. 
Here we just use the fact that for all galaxies in the
sample the whole region in which the bulge dominates the disc is included in the
{\tt SAURON} field.

Contrary to ellipticals, for which absorption line indices are
available for many galaxies, there are very few spiral galaxies with good literature 
measurements. The reason is probably that spirals are more complicated, 
with much more star formation and dust, with
several clearly distinguishable morphological components such as bulge and disc, and more
difficult to observe, given  their generally lower surface brightness. 
While for elliptical galaxies one can obtain a significant amount of useful 
information from broadband
colours, for spiral galaxies these colours are hard to interpret due to the much more
ubiquitous presence of dust.
It is therefore timely and important that more absorption line strength observations of spirals
become available.

We will now briefly summarize the literature on absorption line strengths in spiral galaxies. Note 
that the galaxies for which measurements are available are almost all early-type spirals (type Sbc
or earlier).
Spinrad \& Taylor (1971) already noticed the strong lines (i.e. high metallicity) in the
central regions of M~31, and the complicated spectrum of M~81 with emission and absorption
lines. Bica (1988) fitted spectra of several spiral galaxies to a library of star clusters
and found that young stellar populations occur more often in spirals than in ellipticals.
While calibrated Mg$_2$ indices for large samples of early-type galaxies were already
available in the beginning of the nineties (Faber et al. 1989, Bender et al. 1993),
only in 1996 calibrated Mg$_2$ indices and a number of other
uncalibrated  line indices were published for a sample of S0 and spiral galaxies by
Jablonka, Martin \& Arimoto (1996). They found a reasonably
tight relation  between Mg$_2$ and central velocity dispersion. They also found that the
Mg/Fe ratio in the  centre is higher than solar for bright galaxies, and close to solar for
the faintest galaxies  (M$_R$ $>$ -19). Idiart, de Freitas Pacheco \& Costa (1996) observed
Lick indices in the centres of  early type spirals. They found correlations of the small
number of indices observed with both bulge luminosity and velocity dispersion. Proctor \&
Sansom (2002) published a large number of Lick indices for a sample containing 15 spirals.
They found that bulges are less enhanced in light ($\alpha$-capture) elements and have
lower average age than early-type galaxies. 
A detailed comparison with this sample is given in Section 4. Afanasiev \& Sil'chenko
(2005) show aborption line maps of two early-type spirals in the Leo group: NGC 3623 and NGC
3627, partly using the same {\tt SAURON} data that we present here, but reduced in a 
different way. Gorgas, Jablonka \& Goudfrooij (2007) and Jablonka, Gorgas \&
Goudfrooij (2007) comment on observations of 31
spirals that they observed on the minor axis. They find that their vertical
line strength
gradients are generally negative, and agree very well with the results found for 
elliptical galaxies and S0s from Gonz\'alez \& Gorgas (1995). We conclude
that there is a concensus that Mg/Fe ratios for spirals seem to be somewhat lower than for
ellipticals. The results for the ages of the stellar populations in the centres of spirals
are still under discussion. Fisher et al. (1996) investigated major and minor axis line
strength gradients of S0 galaxies. Interestingly enough, they found much larger gradients
along the minor axis than along the major axis. Along the minor axis, they found negative 
gradients, consistent with stellar population gradients from colours, that are larger than
those in typical elliptical galaxies. Very recently Moorthy \& Holtzman (2006) published a 
large absorption line strength study of long-slit spectra of 38 spirals of type S0-Sbc.  They
separated their sample in red (B--K $>$ 4) and blue bulges. According to them red bulges
of all Hubble types have stellar populations similar to luminous elliptical galaxies. Blue 
bulges consist of metal-poor bulges with low velocity dispersion, and young, metal-rich bulges
that contain all Hubble types and velocity dispersions. Bulges and ellipticals show a continuous
and overlapping sequence in index - $\sigma$ diagrams. Most blue bulges have solar $\alpha$/Fe
ratios. 

The star formation (SF) history of spirals has traditionally been studied using their H$\alpha$
ionised gas emission. Massive SF can be convincingly traced by the accompanying H$\alpha$
emission and is very easily observed with standard telescopes and cameras (Kennicutt 1998). 
H$\alpha$ is mainly produced in the HII regions surrounding massive B and O stars, although
shocks and non-stellar activity can also lead to H$\alpha$ emission. Knapen et al. (2006)
recently studied  the morphology of the H$\alpha$ emission in the circumnuclear regions, 
as well as from the nucleus {\it per se}, using a sample with some prior
evidence for the presence of H$\alpha$. These authors conclude that H$\alpha$ is often found in 
circumnuclear regions such as inner rings, with diameter smaller than 2 kpc. Such
low-luminosity starbursts are found in around one fifth of spiral galaxies (Knapen 2005,
hereafter K05), and are believed to be directly related to  the dynamics of the host galaxy
and its stellar bar (e.g., Buta \& Combes 1996; Knapen 2005). Can this phenomenon also be
seen from the absorption line indices? What is the distribution of young stellar 
populations? 

This question is important when one wants to study the origin of bulges and discs.  At
present, there are a number of theories about the origin of bulges. One is the  theory of
dissipative collapse (Eggen, Lynden-Bell, \& Sandage 1962), where the  bulge assembled from a
primordial proto-galaxy on a short timescale. Such a collapse  is thought to create a bulge
that looks like an elliptical galaxy, also sometimes called a  {\it classical} bulge (e.g.
Carollo 1999). 
However, since bulges live inside discs, some of which are much larger than them, there must
be interactions with them. There are also theories predicting that bulges are made from disc
material. In those so-called {\it secular evolution} scenarios, in which the bulge is formed
by dynamical instabilities of the disc, these instabilities are often bar-driven. Bars might 
lead to radial inflow of
material, buckle and thicken in the vertical direction, and may even be destroyed by
sufficiently massive central mass concentrations (e.g. Combes et al. 1990; Pfenniger and
Norman 1990; Norman et al. 1996). Those processes may occur repetitively, especially in the
presence of gas infall, gradually building a bulge (e.g. Bournaud \& Combes 2002; Mart\'\i
nez-Valpuesta, Shlosman \& Heller 2006). One might possibly be able to distinguish 
between  both models by looking
at the stellar populations (a bulge made from disc material should be younger than a
classical bulge), the morphology (a disc-like bulge should be thinner) or the kinematics (it
should be more supported by rotation). A detailed discussion about the different formation
models, and their comparison with data is given in Kormendy \& Kennicutt (2004). In this
paper we consider what the {\tt SAURON} data, and especially the stellar populations can tell us 
about the formation of spirals.

The paper is structured as follows. In Section~\ref{sec:observations} we  briefly
summarize our observations and data reduction.  In Section 3 we present the
absorption line maps. In Section 4 we discuss central ages, metalicities and abundance ratios,
determined assuming that the stellar populations can be represented by SSPs.
In Section 5 we discuss the central indices, and their
correlations with various relevant galaxy parameters. 
In Section 6 we discuss the relation between the indices and the dynamics, as derived in 
particular from the central velocity dispersion.
In Section 7 we discuss the
implications of these data to our understanding of spiral galaxies, bulges and discs.
Conclusions are given in Section 8.
Finally, in the appendix the individual objects are discussed.  In a forthcoming 
paper, we will discuss line strengths as a function of radius.

\section{Integral-field spectroscopic observations}
\label{sec:observations}

The {\tt SAURON} survey is a study of the
two-dimensional kinematic and stellar population properties of a 
representative sample of 48 early-type galaxies (E + S0) and 
24 early-type spirals.
Details about the definition and properties 
of the survey sample can be found in de Zeeuw et al. (2002) (Paper II).
The observations were carried out using the Integral Field
Spectrograph {\tt SAURON} on the 4.2m William Herschel Telescope of the 
Observatorio del Roque de los Muchachos at La Palma, Spain, and
were performed during 6 runs from 1999 to 2004. 
Paper VII (Falc\'on-Barroso et al. 2006a)
presents the kinematics of gas and stars of the sample of 24 early-type
spirals for which we present the absorption line strength distribution here. 
Most of these galaxies have been classified as Sa in the RC3, although a few
have S0$^+$, S0/a, Sab or Sb classifications. 
In this paper also practical details about the sample of this paper are given,
showing the total integration time, the observing run and the number of
pointings for each object. The sample is summarised in Table 1.
The field of {\tt SAURON} is 33$\arcsec$ $\times$ 41$\arcsec$,
with a spatial sampling of 0\farcs94 $\times$ 0\farcs94. Although for 
most galaxies we only observed one position, we performed two pointings for three of 
the largest galaxies. Details about the instrument are given in Paper I (Bacon
et al. 2001). The data and maps presented in this paper will be made
available via the {\tt SAURON} WEB page {\tt http://www.strw.leidenuniv.nl/sauron/}.

The data reduction was performed with the \Xsauron\ package, providing cubes of
spectra covering a narrow wavelength range 4800 -- 5300 \AA\ at a resolution of
4.2 \AA (FWHM) and a sampling of 1.1 \AA/pix. In this wavelength range one finds
the Lick indices \hb\, \fe\ , \mgb\ and Fe 5270, and the emission lines \hb\,
[OIII] at 4959 and 5007 \AA\ and the [NI] doublet at 5199 \AA. More details about the data
reduction are given in Paper VII. In this paper the Fe 5270 index is not used,
since its maps cover only slightly more than half the field of the other indices.

To allow for a good calibration of the line indices, also from run to run, we observed during
each run a number of stars covering a broad range of spectral types. Specifically, we
included stars from the Lick stellar library catalogue (Worthey et al. 1994) and the MILES library
(S\'anchez-Bl\'azquez et al. 2006) in order to
calibrate our line strength measurements to the flux-calibrated Lick/IDS system 
and its associated models
(e.g. Worthey 1994, Vazdekis et al. 1996). Spectrophotometric standard stars were 
observed to calibrate the response function of the system (Paper VI), where 
the observations of stars and
elliptical galaxies are compared with observations from resp. Worthey et al. (1994) and
Trager et al. (1998), showing that these measurements can be reproduced within the errors of
the Lick system. In Paper II a detailed comparison with the literature is made for two
ellipticals: NGC 3384 and NGC 5813, showing good agreement.

The spectra were fitted with the stellar population models of Vazdekis
(1999), allowing us to separate the emission lines from the absorption
line spectrum (for details about this procedure see Papers V and
VII). On the cleaned spectra we obtained the line indices (see Paper
VI for details on how this was done). %
The simulations in Appendix A of Paper V show that the accuracy in
recovering the emission-line fluxes does not depend on the strength of
the emission, as quantified by the A/N ratio between the line
amplitude and the noise level in the stellar continuum. This is
because the errors in matching the line amplitudes do not depend on
the A/N ratio but only on the amount of noise in the
continuum. Considering only statistical fluctuations, the uncertanties
in the fluxes of the emission lines will therefore increase with
increasing S/N in the stellar continuum but the uncertainties in the
equivalent width of the lines will decrease.

For the worst S/N=60, the typical uncertainties in the line fluxes in
Appendix A of Paper V (see Fig. A1 of that paper) translates in errors
in the emission line equivalenth width of $\sim 0.08$\AA, which will
correspond to similar errors in the \hb,\fe\ and \mgb\ line indices.
The fact that the spectra are packed so close together implies that 
neighbouring spectra, at wavelengths by a few hundred angstroms, will affect
the absorption lines discussed here. Although our reduction programs have 
been optimized to get rid of this contaminating emission, it is unavoidable 
that some effects can not be removed. This is the case in particular when 
the galaxy has strong emission lines. Other, smaller errors arise because 
of template errors between the galaxy and the input spectra of the stellar library
of Vazdekis (1999). Considering these points, we  presume throughout the rest of 
this paper that the
uncertainty in the data points amounts to more conservative values of 
0.2 \AA\ for \mgb\ and \hb\ and 0.3 \AA\ for \fe. 

A few of the galaxies (NGC 4369, NGC 4383) have emission lines that
are stronger by a factor of about 100, relative to the absorption
lines, than the elliptical or lenticular of paper V with the strongest
emission lines. We have done some extra simulations, similar to the
ones of Appendix A of Paper V, with A/N now ranging up to 100. Since
the results are such that the errors in the line indices \hb,\fe\ and
\mgb\ do not increase noticibly, we use the same errors in this paper.
As a test we also used the MILES stellar population models
(S\'anchez-Blazquez et al. 2006) to separate absorption and
emission. The resulting line strengths are the same as using the
Vazdekis (1999) models within the errors.

\section{Absorption-line strength maps}

\subsection{The data}

Figure~1 present maps of the absorption
line strengths of the 24 objects, ordered by increasing NGC number. 
In the first row, we show the measured two-dimensional line strength
distributions of H$\beta$, Fe5015 and \mgb\ . In the second row, the
total intensity map reconstructed from the {\tt SAURON} spectra is
followed by the age, metallicity and Mg/Fe overabundance maps derived
from single stellar population (SSP) models (see Section 3.2). 
The relative directions of North and
East are indicated by the orientation arrow next to the galaxy title 
(the orientation is identical to Paper~VII). The maximum and minimum of the
plotting range is given in the tab attached to each parameter map, and the
colour bar indicates the colour table used. 

\begin{table}
\begin{tabular}{lllcclccc}
Galaxy  &   Type & SF Type & $\epsilon$ & M$_B$ & Activity & W$_{20}^c$ &  $\sigma_{cen}/\sigma_{max}$  &
$\sqrt{\sigma_{max}^2-\sigma_{cen}^2}$ \\
(1) & (2) & (3) & (4) & (5) & (6) & (7) & (8) & (9) \\
NGC & RC3 & RC3 & RC3 & Paper II & H97 & RC3 & Paper VII & Paper VII \\
\hline
1056 & Sa:         & CR &  0.44  &  -19.44     & Sy2$^*$    & 311  &     0.83   &                 52    \\
2273 & SBa(r):     & CS &  0.25  &  -20.21     & Sy2  & 513    &  0.83   &			  70    \\ 
2844 & Sa(r):      & R &  0.55  &  -18.38     & HII$^*$     & 334   &	 0.94   &		  34    \\
3623 & SABa(rs)    & N (D) &  0.76  &  -20.82     & L2::    & 520 & 		 0.80   &	  100   \\
4220 & S0$^+$(r)   & LR &  0.61  &  -19.01     & T2   & --   &		 0.99   &		  16    \\
4235 & Sa(s)sp     & N (D) &  0.78  &  -19.20     & Sy1.2   & 335   & 	 0.84   &		  90	\\
4245 & SB0/a(r)    & R &  0.17  &  -18.72     & HII   & --   &  	   0.92   &		  36	\\ 
4274 & (R)SBab(r)  & R &  0.63  &  -20.08     & HII    & 481 &  	   0.68   &		  100	\\ 
4293 & (R)SB0/a(s) & CS &  0.48  &  -20.37     & L2 & 424   &	  0.92   &			  42	\\
4314 & SBa(rs)     & R &  0.05  &  -19.55     & L2   & -- &		 0.93   &		  43	\\
4369 & (R)Sa(rs)   & CR &  0.01  &  -18.96     & HII   & --   &  	0.99   &		  10	\\
4383 & Sa pec      & CR &  0.49  &  -18.93     & HII$^*$    & 237   &	1.00   &		  0 	\\ 
4405 & S0/a(rs)    & CR &  0.34  &  -18.54     & HII   & 187   & 	0.91   &		  25	\\
4425 & SB0$^+$:sp  & N	   &  0.64  &  -18.83	  &	   & --  &	       0.93   & 	  28	\\
4596 & SB0$^+$(r)  & N (D) &  0.13  &  -19.94	  & L2::   & -- &	       0.92   & 	  59	\\
4698 & Sab(s)      & N (D) &  0.31  &  -20.05	  & Sy2     & 544 &	       1.00   & 	  10	\\
4772 & Sa(s)       & N (D) &  0.42  &  -19.56	  & L1.9   & 531 &	       0.99   & 	  15	\\ 
5448 & (R)SABa(r)  & CS &  0.48  &  -20.78     & L2 & 464     &   0.86   &			  65	\\
5475 & Sa sp       & N  &  0.74  &  -19.39     &	 & -- & 	    0.99   &		  14	\\
5636 & SAB(r)0+    & N  &  0.32  &  -18.42     &	 & 430 &	    0.99   &		   8	\\
5689 & SB0$^0$(s)  & N (D)	  &  0.71  &  -20.32	 &	  & 381  &	      1.00   &	  0 	\\
5953 & Sa:pec      & R  &  0.26  &  -19.61     & L1/Sy2$^*$  & 363&	    0.66   &		  81	\\ 
6501 & S0$^+$      & N  &  0.10  &  -20.38     & L2::		 & 503 &    0.99   &		  27	\\
7742 & Sb(r)       & R  &  0.05  &  -19.76     & T2/L2 & 267&		    1.00   &		  0 	\\
\hline												    
\end{tabular}

{\bf Notes to Table 1:} Some global parameters for our galaxies. (1): NGC number; 
(2) Morphological type (from RC3, de Vaucouleurs et al. 1991); (3) Morphology of 
the central star formation region  (see text: CR = central region; CS = central starburst; R = ring, LR = large
ring, N = no signs of significant recent star formation. (D) indicates the presence of
a central disc) (4): Ellipticity 1-b/a (RC3); (5) Absolute blue magnitude (Paper II), 
(6) Activity class (Ho et al. 1997). Asteriscs indicate that the galaxy is not included in H97 and that the
classification is from NED. Column (7) gives the inclination corrected 
HI velocity width at 20\% of the peak (from NED), in km/s. In column (8) the central stellar velocity 
dispersion is given divided by the maximum velocity dispersion in the {\tt SAURON} field, and 
in column (9) a similar quantity, in km/s, not scaled by the central velocity dispersion.

\end{table}

\onecolumn
\begin{figure}
\caption[]{Absorption line strength maps of the 24 Sa galaxies in the
{\tt SAURON} representative sample. The {\tt SAURON} spectra have been spatially binned
to a minimum signal-to-noise of 60 by means of the Voronoi two-dimensional
binning algorithm of Cappellari \& Copin (2003). For each
galaxy its orientation is indicated by the arrow behind its NGC
number, pointing to the North and the associated dash to the
East. The corresponding position angle of the vertical (upward) axis is  provided in
Table~2 of Paper VII. Shown are (from left to right): 
line indices \hb\ ,\fe\  and \mgb.
Second row: Reconstructed intensity, logarithmic Age, Metallicity 
(log Z/Z$_\odot$) 
and [$\alpha$/Fe] (for details about how these parameters were 
obtained see text). The reconstructed intensity is overlayed in 
white contours on the maps.{\bf Figure 1a available as separate jpg-file} }
\label{fig:ls1}
\end{figure}
\addtocounter{figure}{-1}

\begin{figure}
\caption[]{b - continued. Figure 1b available as separate jpg-file}
\label{fig:ls2}
\end{figure}
\addtocounter{figure}{-1}

\begin{figure}
\caption[]{c - continued. Figure 1c available as separate jpg-file}
\label{fig:ls3}
\end{figure}
\addtocounter{figure}{-1}

\begin{figure}
\caption[]{d - continued. Figure 1d available as separate jpg-file}
\label{fig:ls4}
\end{figure}
\addtocounter{figure}{-1}

\begin{figure}
\caption[]{e - continued. Figure 1e available as separate jpg-file}
\label{fig:ls5}
\end{figure}
\addtocounter{figure}{-1}

\begin{figure}
\caption[]{f - continued. Figure 1f available as separate jpg-file}
\label{fig:ls6}
\end{figure}
\addtocounter{figure}{-1}

\begin{figure}
\caption[]{g - continued. Figure 1g available as separate jpg-file}
\label{fig:ls6}
\end{figure}
\addtocounter{figure}{-1}

\begin{figure}
\caption[]{h - continued. Figure 1h available as separate jpg-file}
\label{fig:ls6}
\end{figure}
\addtocounter{figure}{-1}

\twocolumn

\subsection{Stellar Population Structures in the {\tt SAURON} maps}

In our sample of 24, we distinguish two kinds of galaxies: those with smooth
line strength gradients, and those with features for which \mgb\ and \fe\ are
considerably
lower than the surrounding areas and \hb\ higher, i.e. areas with
younger stellar populations. These younger stellar populations are generally
found either in rings, or everywhere in the central regions; the presence of
young stars always comes with the presence of dust. 

Starting with the second category: we find rings of younger stars in NGC 2844, 4220,
4245, 4274, 4314, 5953 and 7742. The appearance of the line strength maps in
e.g. NGC 2844 and 4220 is different from those in e.g. NGC 4314, because of the
effects of inclination.
  
\begin{table}
\begin{tabular}{l|cccc|cccc}
Galaxy  &  \multicolumn{2}{c}{H$\beta$} &  \multicolumn{2}{c}{Mg b} & 
\multicolumn{2}{c}{Fe 5015} & \\
    & 1.2$''$  & 10$''$ &   1.2$''$ & 10$''$  &   1.2$''$ & 10$''$ \\
\hline
NGC 1056&  3.16   & 2.85   &    1.63  &	  1.55 &    2.65    &   2.25   \\
NGC 2273&  3.33   & 2.71   &    2.07  &	  2.88 &    2.88    &   4.30  \\
NGC 2844&  2.13   & 2.54   &    2.72  &	  2.01 &    4.27    &   3.25  \\
NGC 3623&  1.78   & 1.70   &    4.13  &	  3.96 &    5.80    &   5.00  \\
NGC 4220&  2.53   & 2.52   &    3.05  &	  2.81 &    5.67    &   5.00  \\
NGC 4235&  1.27   & 1.84   &    3.02  &	  3.42 &    3.75    &   4.74  \\
NGC 4245&  2.08   & 2.12   &    3.80  &	  3.27 &    5.66    &   4.57  \\
NGC 4274&  2.24   & 2.06   &    3.53  &	  3.29 &    5.96    &   4.58  \\
NGC 4293&  3.64   & 2.59   &    2.33  &	  2.79 &    4.68    &   4.97  \\
NGC 4314&  1.81   & 2.29   &    3.89  &	  2.87 &    5.50    &   4.16  \\
NGC 4369&  4.37   & 3.52   &    1.20  &	  1.46 &    3.03    &   2.47  \\
NGC 4383&  2.69   & 3.11   &    0.89  &	  1.27 &    0.31    &   1.25  \\
NGC 4405&  3.54   & 3.32   &    1.72  &	  1.78 &    3.56    &   3.05  \\
NGC 4425&  2.00   & 1.96   &    3.47  &	  3.17 &    5.44    &   4.51  \\
NGC 4596&  1.86   & 1.71   &    4.20  &	  3.90 &    5.79    &   4.79  \\
NGC 4698&  1.55   & 1.66   &    4.08  &	  3.56 &    5.28    &   4.48  \\
NGC 4772&  1.52   & 1.50   &    4.23  &	  3.87 &    4.59    &   4.32  \\
NGC 5448&  2.77   & 2.10   &    2.69  &	  3.27 &    4.66    &   4.38  \\
NGC 5475&  2.40   & 2.20   &    3.33  &	  3.14 &    5.54    &   4.81  \\
NGC 5636&  2.56   & 2.34   &    2.02  &	  2.23 &    3.63    &   3.20  \\
NGC 5689&  2.05   & 1.98   &    3.73  &	  3.28 &    5.83    &   4.86  \\
NGC 5953&  2.28   & 3.24   &    2.04  &	  1.45 &    3.14    &   2.42  \\
NGC 6501&  1.37   & 1.55   &    4.94  &	  4.11 &    5.86    &   4.99  \\
NGC 7742&  2.41   & 2.90   &    3.05  &	  2.28 &     4.63    &   3.56  \\
\hline
\end{tabular}

{\bf Notes to Table 2:} Here we present central line indices in apertures of
 1.2$\arcsec$ and 10$\arcsec$ radius. The numbers are given in the 
 Lick system, i.e., corrections of -0.13 \AA\ and 0.28 \AA\ resp. have been added to 
 H$\beta$ and \fe\ (Paper VI).

\end{table}

In the same category we find a class of galaxies with young stellar populations
in the central regions. Here we distinguish again two subgroups. The first group,
consisting of NGC 2273, 4293, and 5448, contains a compact central region with
diameter of about 10 arcsec or $\sim$ 1 kpc, inside an otherwise old stellar
population. The other group contains galaxies for which the stellar
populations are young across the whole {\tt SAURON} field (NGC 1056, 4369,
4383, 4405). These are amongst the faintest galaxies of our sample, but HST images 
show that there is very little star formation further out in the disc. If there had been, 
they probably would not have
been classified as Sa in the RC3. The [OIII]/H$\beta$ emission line ratio in
these objects is generally low, indicating the presence of star formation (Paper V).  
For this category the presence of young
stars is always accompanied by large amounts of dust (see the unsharp masked
images from HST and MDM in Paper VII). Seven galaxies in total (29 $\pm$ 9 \%)
belong to this  class. At higher spatial resolution some of these might
change to the  previous category.

The remaining galaxies, NGC 3623, 4235, 4425, 4596, 4698, 4772, 5475, 5636, 5689
and 6501 show generally smooth line strength maps, similar to elliptical and
lenticular galaxies. In six of those ten there are clear indications of dust in
the central regions.

The young stellar populations that we see in the absorption line strength maps
are always detected in the H$\beta$ emission line maps (shown in Paper VII).
Features in the line strength maps are often, but not always seen in the
stellar and gaseous kinematics. In galaxies with a
ring at high inclination (e.g. NGC 4274) we see that the ring rotates rapidly,
and is associated with gas, dust, and regions of young stellar populations.
Here one can clearly see that the young stellar populations are confined to a
flat, disc-like region. Another correlation between the presence of young population and the
kinematics can be found in galaxies with
central young stellar populations. For example,  in NGC 2273 we find a thin
disc in the stellar kinematics in the inner regions. A region with the same size as this disc
along the major axis, but more extended on the minor axis shows evidence of
younger stellar populations. In the galaxies
with rings, sometimes stellar discs are seen inside the ring (e.g. NGC 4245,
4314). In NGC 4293 we see central young populations associated with abnormally
low gas velocities, indicating possibly ionised gas in
outflow. In NGC 4698 we clearly see a central peak in both the \mgb\ and \fe\ maps,
indicating a high metallicity.
This galaxy has a central stellar disc rotating perpendicular to the rest of the 
galaxy (Pizzella et al. 2002, Sarzi 2000). 
This combination of a kinematically decoupled core and an enhanced central
metallicity is also seen in several central discs in elliptical and S0
galaxies (NGC 3414, 3608, 4458, 4621, 5198, 5813, 5831, 5982 and 7332; Paper VI, and NGC 4365
(Davies et al. 2001)). NGC 5953, part of an interacting pair, shows a ring of
young stellar populations. The stars inside this ring rotate perpendicular to
those outside of it. In Paper VII we suggest that we are seeing
here a kinematically decoupled core being formed. Finally, in NGC 7742 the data
in Paper VII show that the ionised gas is not only counter-rotating to the
stars inside the ring (as was shown in Paper II), but also outside of it. In
the ring itself, dominated by young stellar populations, the rotation velocity
is lower than immediately inside and outside of it. 

How do these Sa galaxies relate to earlier-type galaxies, e.g. S0s? 
In Paper VI we mention in Section 5.1 that in
NGC 3032, 3156, 4150 and 4382 central depressions in \mgb\ are found,
corresponding to regions of enhanced H$\beta$. All 4 galaxies are lenticulars.
We can compare them to the spirals with central starbursts. 
In the S0 galaxies NGC 524 and NGC 3608 there is some evidence for rings,
associated with younger stellar populations (Paper V, Paper VI), although the
amount of light from the young stars is much smaller than in the ring galaxies
in this paper. Also, this paper contains a case of a galaxy with young stars in
a ring (NGC 4526), similar to NGC 4274. 

\section{SSP Ages, Metallicities and Abundance Ratios}

In the way we described in Paper VI and in McDermid et al. (2006, Paper VIII) 
we determined ages, metallicities
and abundance ratios in each bin, assuming that the stellar populations there could
be represented by a single-age, single metallicity stellar population. In practice,
we determined the SSP for which the line strengths \fe\ , \hb\ and \mgb\ fitted best
in the $\chi^2$ sense. These maps are shown in the second row of Figure 1. Although we
know that it is a great over-simplification to represent the stellar populations
(even locally) of a galaxy by a SSP (e.g., Allard et al. 2006), in some, especially elliptical
galaxies (e.g. NGC 5128, Rejkuba et al. 2005) the locally  averaged
metallicity and age do not vary very much across the galaxy, so that the errors that
one makes when representing the local stellar population by an SSP are the same
everywhere. For that reason by far the large majority of papers dealing with stellar
populations in ellipticals treats these as SSPs. In Paper VI one can see that among
the S0 galaxies some have stellar populations with  different ages, such as NGC 3032
in the very nucleus (see also OASIS data of this galaxy in 
Paper VIII). In this galaxy a representation in terms of SSPs removes a considerable
amount of information and can lead to wrong results. Such young stellar populations,
however, are rare in the survey of elliptical galaxies and S0s (see Paper VI).  The
line strength maps of Fig. 1 show that for Sa-galaxies the situation is different.
Several galaxies show features in their age distribution, indicating younger stellar populations. 
Apart from the absorption line maps, the
emission line maps also show that  H$\beta$ is
sometimes strong, indicating stars of around 10$^8$ years. 
Some galaxies have absorption line strength maps without features,
just like elliptical galaxies. Others are very different. For the  former objects the
SSP approach might give results that are close to reality. For the latter objects,
where the line strength maps show features, one should just consider the age,
metallicity and alpha-enhancement maps as tools, and interpret them with the caveats
given here.  

In some objects an unconstrained SSP-fit gives rather inappropriate results. This
is illustrated in Allard et al. (2006) for the star formation ring in NGC 4321.
Allard et al. (their Fig. 12) show that in the ring \mgb\ , \fe\ and \hb\ are such that the
stellar populations there have to consist at least of two components: a young one,
and an old, metal-rich stellar population. If one forces only one Single Stellar
Population, it will have low metallicity and old age. Since metallicities of HII
regions in this galaxy are likely to be higher than solar (Zaritsky et al. 1994)
the SSP solution looks clearly wrong. This seems to be the case in particular for 
NGC 4314, 4369, 4383, 5953, and 7742. For these especially one has to understand 
the limitations of the ages obtained here.

\section{Central stellar populations}
\label{sec:central}

In this section we analyse the central line strength indices, calculated in the central
aperture with a radius of 1.2$\arcsec$, and their corresponding
SSP-metallicities, ages and $\alpha$/Fe ratios, and discuss their dependence on 
other galaxy parameters.

\subsection{Index-index relations}

In Fig.~2 and 3 we show two different index-index diagrams.  The first (Fig. 2)
is a metal line indicator (MgFe50) against an age indicator (H$\beta$). 
The numbers used in these figures are given in Table 2. 
Indices are measured in a central aperture with radius 1.2$\arcsec$.
MgFe50 is defined to be
$\sqrt{{\rm Mg} b \times {\rm Fe 5015}}$ (Kuntschner 2000), and has been shown to be
a good metallicity indicator, relatively unaffected by the effects of the
overabundance of Mg with respect to Fe. In red squares are indicated the elliptical and
lenticular galaxies of Paper VI (at r$_e$/8), and in
filled blue circles the central apertures of the  Sa galaxies of this paper. 
The choice of aperture is not arbitrary. If we would take the 
same aperture of r$_e$/8 as in Paper VI we would have to make a choice of 
either performing a bulge-disc decomposition and taking r$_e$/8 of the bulge,
or taking r$_e$/8 of the whole galaxy. For the spiral galaxies we have taken the 
approach that we would concentrate our efforts on the inner regions, and therefore
only observed one {\tt SAURON} field per galaxy, as opposed to many of the early-type galaxies 
of paper VI, for which 2 or 3 fields were observed. Consequently r$_e$/8 of the bulge would 
be the natural choice for the inner aperture.
For bulges, the effective radius generally would be smaller than 10$\arcsec$ (Andredakis
et al. 1995), so 
r$_e$/8, corrected for the effects of the seeing, would be comparable to 1.2$\arcsec$. 

From Fig.~2 one can see that
there is a smooth transition between E and S0 galaxies on one hand and Sa galaxies on the
other. Most early-type galaxies can be
found in the lower right part of the diagram, while the Sa galaxies have lower
MgFe50 and higher  H$\beta$ values. Added to the diagram is a grid of SSP
models by Thomas, Maraston \& Bender (2003). Most early-type galaxies can
be interpreted as having old, metal rich stellar populations (similar to
galaxies in the Fornax cluster, see Kuntschner 2000), but some are clearly
younger in their central regions.  The Sa galaxies apparently have a large
range in age, and have metallicities generally lower than the ellipticals, if we 
assume that we are dealing with SSPs.  One
of the galaxies, NGC 4383, has a much lower metallicity than the other
galaxies, or is dominated by very young stellar populations. NGC 4235 lies slightly below the grid. The line strength maps of
Figure 1 indicate a central dip in H$\beta$, which makes one suspect that this
Seyfert 1 galaxy has  some non-thermal emission in its very centre. \\

In Figure 3 we investigate the [Mg/Fe] overabundance in spirals.  Since some
elliptical galaxies are known to be over-abundant in $\alpha$-elements,
compared to the Sun (Peletier 1989, Worthey et al. 1992), we have plotted in the bottom panel
models by Thomas et al. (2003) with  $[\alpha/Fe]$=0.5 (dotted lines) and 0 (solid lines).
There is a general tendency for
Sa galaxies with a given velocity dispersion to have the same abundance ratio [Mg/Fe]
(in the center) as elliptical galaxies. 
The fact that the abundance ratios of the two types of galaxies are
the same shows that the star formation history in the centers of these galaxies has been
very similar.
At low velocity dispersion the scatter for spirals
is large, but this is most likely due to the fact that these galaxies consist of a mix of
young and old stellar populations, implying that our $\chi^2$ method to derive the abundance
ratio breaks down here. One can illustrate this as follows: 4 of the spirals have been indicated 
with blue-yellow open symbols. These four are the objects that deviate most from 
the early-type galaxies and show apparently high [Mg/Fe] for low central velocity dispersion.
The objects are NGC 1056, 2273, 4383 and 5953. Inspection of their line strength maps shows that 
all 4 have large central H$\beta$ values, indicating objects in which a significant fraction 
of the light comes from young ($\sim$ 1 Gyr) stellar populations. These objects most likely 
cannot be fitted with SSP models, so for those we cannot use the \mgb\ - \fe\ diagram to 
derive their [Mg/Fe] abundance ratio.

Following the currently most popular explanation of nucleosynthesis
models Mg predominantly comes from Supernovae type II, while Fe mainly comes from type
Ia (Worthey et al. 1992, Weiss, Matteucci \& Peletier 1995). The star formation history
of elliptical galaxies is thought to be such that most of the stars are formed in the 
first Gyr, while this timescale for spirals is supposed to be much longer. Note however
that there are several modes of star formation in spirals. If star formation is
quiescent, which is generaly the case for late type spirals, solar Mg/Fe ratios would be
expected. Indeed, some of the galaxies seem to have [Mg/Fe]=0. If star formation occurs
in bursts, which is clearly happening often as well, star formation timescales will be
short, since the gas will be exhausted, and Mg/Fe could climb to larger values. For
fainter galaxies, bursts are relatively more important (see above), so that the scatter
in Mg/Fe is also expected to be larger. 
The fact that the Mg/Fe ratios for massive galaxies are large shows that the stars must
have formed from massive stars, in strong bursts, consistent with H$\alpha$ measurements.
For fainter galaxies star formation must have happened more slowly.


\addtocounter{figure}{1}
\begin{figure}
\begin{center} 
  \includegraphics[draft=false,scale=0.44,trim=0cm 0.cm
    0cm 0cm]{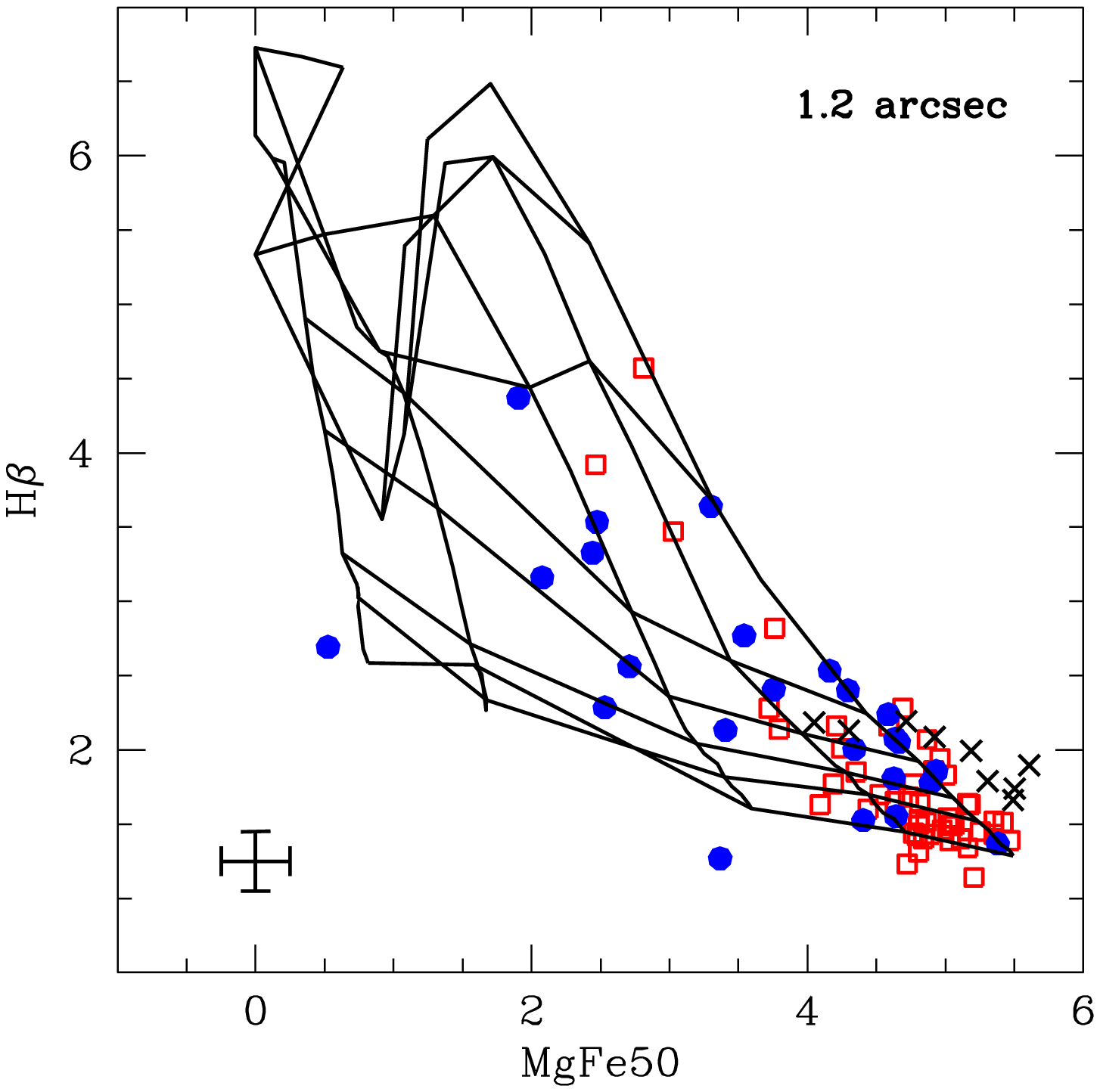} 
\end{center} 
\caption[]{Index-index diagram showing the central line indices H$\beta$ and MgFe50 (in \AA) in circular apertures with
radius 1.2$''$ for the Sa galaxies (in blue,  with representative error bars), together with  integrated indices within r$_e$/8 ( 
red) for the early-type
galaxies of paper VI (in red) , and  with SSP stellar population models by Thomas et al. (2003).  Shown are models with
[Mg/Fe]=0.  In the models metallicity goes up from left to right (from Z=-2.25 to Z=0.35) and age goes up from top to 
bottom (from 0.1 to 15 Gyr). The black crosses
are the central values for the 10 spiral galaxies of Proctor \& Sansom (2002), for which they claim to have reliable
corrections for H$\beta$ emission.  }
\label{fig:indextype}
\end{figure}

\begin{figure}
\begin{center} 
  \includegraphics[draft=false,scale=0.44,trim=0cm 0.cm
    0cm 0cm]{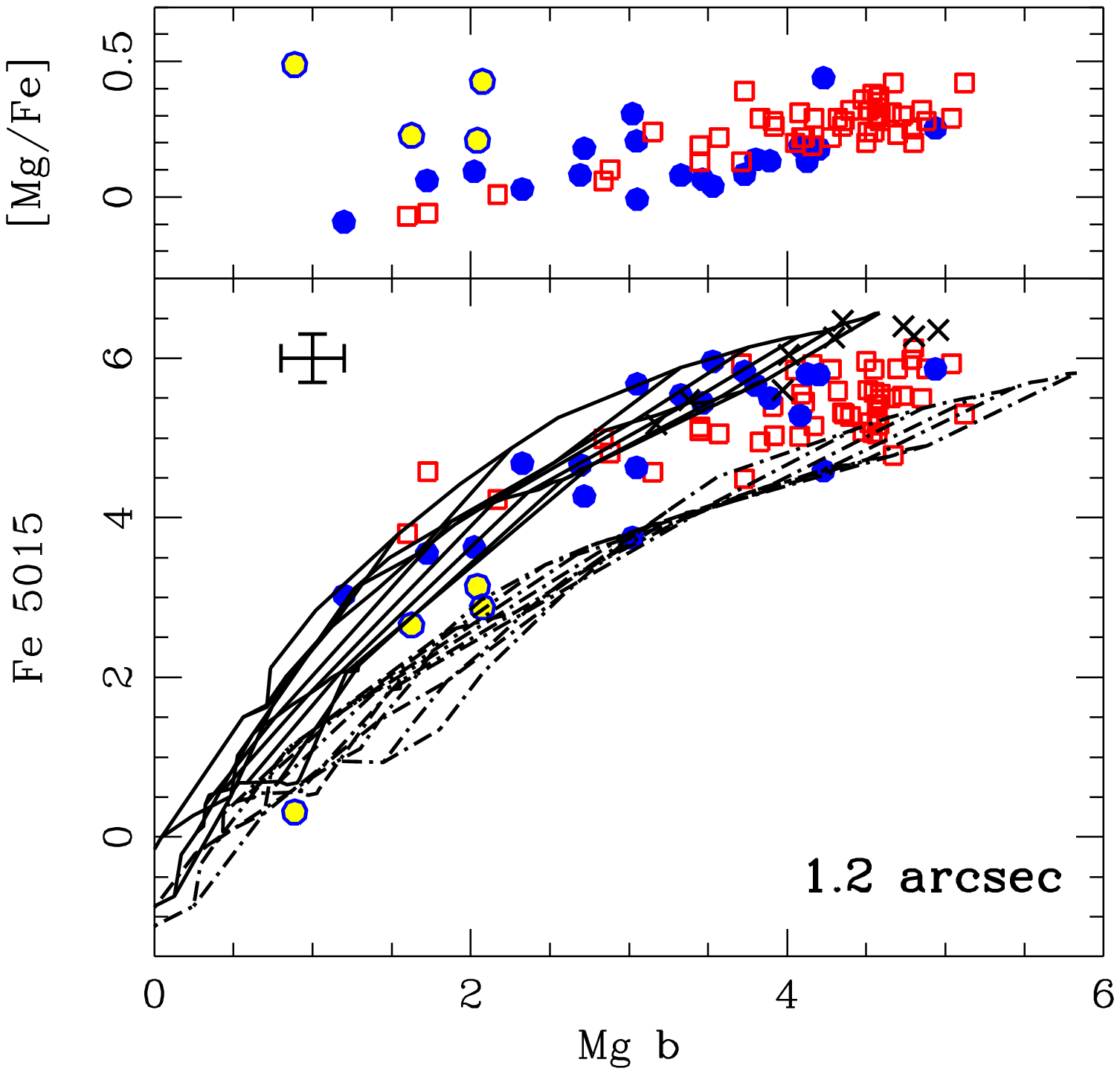} 
\end{center} 
\caption[]{{\bf Bottom}:Index-index diagram showing central line indices \mgb\ and
\fe\  (in \AA) in circular apertures with radius 1.2$''$  for the Sa galaxies (in
blue,  with representative error bars),  together with integrated indices within r$_e$/8 for the early-type galaxies of
paper VI  (in red), and  with SSP stellar population models by Thomas et al. (2003). 
Here models are plotted with [Mg/Fe] = 0 (solid lines) and 0.5 (dashed-dotted), ranging
in metallicity from Z=-2.25 to Z=0.35 and in age from t=0.1 to t=15 Gyr.
The
models with the highest metallicity (Z=0.35) and age (15 Gyr) have the highest \mgb\
and \fe\ values. The black crosses are the central values for the 10 spiral
galaxies of Proctor \& Sansom (2002), for which they claim to have reliable corrections
for H$\beta$ emission. {\bf Top:} [Mg/Fe] ratios calculated from the $\chi^2$ code
(Paper VIII), calculated assuming the stellar populations can be represented by an
SSP. The four blue points filled in in yellow are NGC 1056, 2273, 4383 and 5953, objects 
with large central H$\beta$ values, that most likely cannot be fitted with SSP models.
}
\label{fig:indextype}
\end{figure}



\subsection{Comparison with the literature}

Very few papers have presented absorption line strengths for spiral galaxies.
One of them is the paper of Proctor \& Sansom (2002), which contains data for
15 spirals (type S0a-Sb). Their central data is also shown in Figs. 2 and 3. 
Not shown are the galaxies for which  Proctor \& Sansom
claim that their H$\beta$ absorption line is unreliable because of
uncertainties in removing the emission. In general it seems as if the sample of
Proctor \& Sansom has an offset in Fe 5015 and in H$\beta$, as compared to
ours. There is one galaxy in common: NGC~3623, for which their measurements are
quite different: \hb\ (PS) = 1.664 \AA\ $\pm$ 0.080; $\Delta({\rm H}\beta)$ (PS -- 
{\tt SAURON}) = --0.245 \AA;
\mgb\ (PS) = 4.803 \AA\ $\pm$ 0.065; $\Delta({\rm Mg} b)$ (PS - {\tt SAURON}) = 0.676 \AA; 
\fe\ (PS) = 6.278 \AA\
$\pm$ 0.165; $\Delta({\rm Fe} 5015)$ (PS - {\tt SAURON}) = 0.756 \AA. 
It is unlikely that the difference in \hb\
is caused by errors in removing the H$\beta$ emission, since this galaxy
contains very little emission (see Paper VII). The offsets in NGC~3623 are typical
for the offsets that we see for the  sample as a whole. We do see that the
range in MgFe50 covered by the galaxies of the sample of Proctor \& Sansom
(2002) is much smaller than in our {\tt SAURON} galaxies. As a result, Proctor \&
Sansom find fewer galaxies with recent star formation than we do. There is one
galaxy in common with Proctor et al. (2000), NGC 5689. Here the comparison for the 
central aperture is better: ($\Delta({\rm H}\beta)$ (PS -- {\tt SAURON}) = -- 0.07 \AA, 
$\Delta({\rm Mg} b)$ = 0.30 \AA, and $\Delta({\rm Fe} 5015)$ = -- 0.09 \AA. 
The fact that we don't see any offset between the sample of E and S0's of Paper VI and the
current sample of Sa's, and the good agreement between Paper VI and the literature make us believe
that the data of Proctor \& Sansom might be subject to a systematic offset.

A paper with line indices for a large number of galaxies is Moorthy \& Holtzman (2006). 
Although there is only one galaxy in common, NGC 5689, the behaviour of the galaxies in the 
index-index diagram [MgFe]' vs. \hb\ is similar to our galaxies in the MgFe50 vs. \hb\ diagram.

\subsection{Relation with galaxy morphology}

In Fig.~4 we show the central \mgb\ and H$\beta$ line strength as a function of
morphological type (T-type from the RC3). Here we see the same trends as in Figure
2  (overall \mgb\  decreases as a function of type, H$\beta$ increases, \fe\ (not
shown) decreases, with scatter increasing towards later types),  but what can be
seen very well here is that the scatter in line indices from galaxy to galaxy
becomes large. While for elliptical galaxies line indices occupy a very small range
in equivalent width or magnitude (e.g. Schweizer et al. 1990), the range becomes
larger for S0 galaxies, and this trend increases for later-type galaxies. A commonly
used explanation for this trend is that galaxy populations consist of multiple
bursts (see e.g. Trager et al. 2000, Schweizer \& Seitzer 1992).  A burst of star
formation causes a luminous stellar population with (after about 10$^8$ years) high
Balmer indices, which slowly become weaker with time. 
It is thought that in the nearby Universe  these bursts occur much
more frequently in S0 galaxies than in elliptical galaxies, causing the larger
scatter in the former. Here we show that the same is the case in the central regions
of Sa galaxies. It is very important to realise that we are looking at
luminosity-weighted indices. While most of the mass might be old, a young
population, which always has a very low stellar $M/L$ ratio, could still dominate
the light. Note that both the \mgb\ -- type and H$\beta$ -- type diagram show
envelopes: galaxies have a maximum Mg b and a minimum H$\beta$. These envelopes
represent old, metal rich stellar populations.

\begin{figure}
\begin{center} 
  \includegraphics[draft=false,scale=0.44,trim=0cm 0.cm
    0cm 0cm]{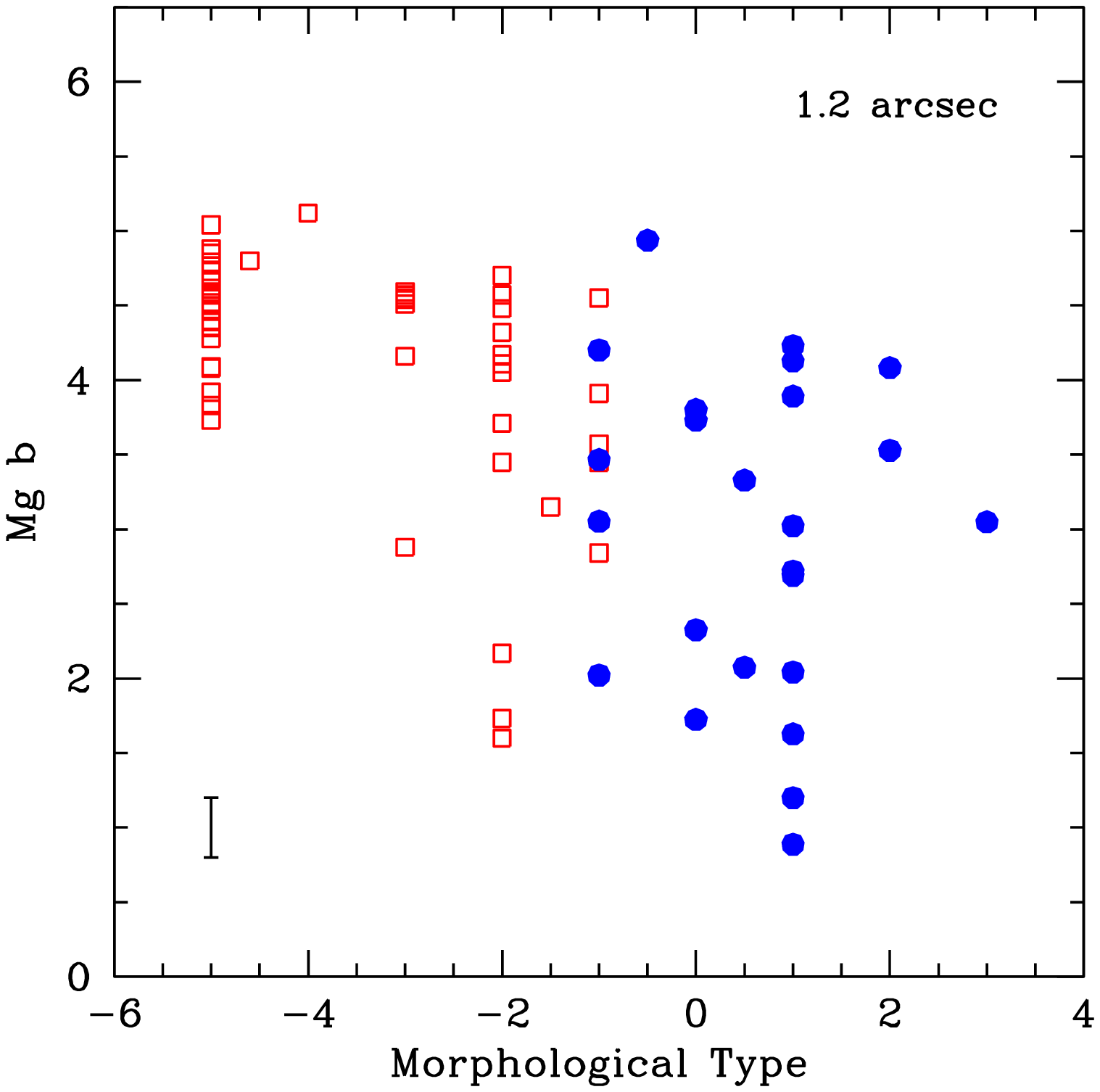} 
  \includegraphics[draft=false,scale=0.44,trim=0cm 0.cm
    0cm 0cm]{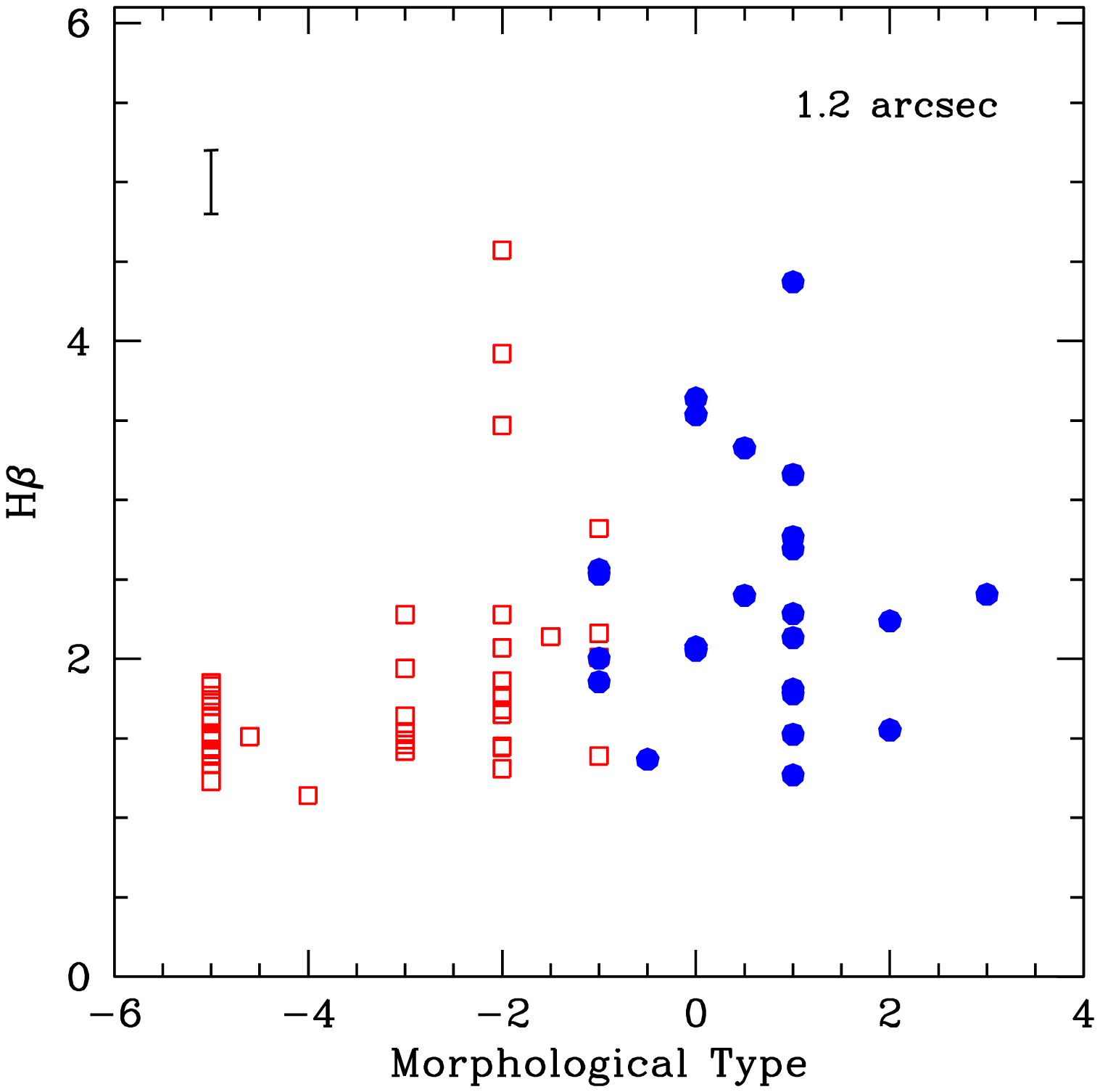} 
\end{center} 
\caption[]{Central indices (in \AA) in an aperture of radius 1.2\arcsec\ as a function of
morphological type. In red is shown the
sample of elliptical galaxies presented in Paper VI. Morphological t-types from
the RC3 (de Vaucouleurs et al. 1991). }
\label{fig:indextype}
\end{figure}

\section{Linking stellar populations with galaxy dynamics}

\subsection{Index -- $\sigma$ relations}

Early-type galaxies show a tight Mg$_2$ -- velocity dispersion relation (Terlevich et al.
1981, Guzm\'an et al. 1992, J\o rgensen et al. 1996). This is one of the important
relations linking galaxy mass with
their stellar populations. Deviations from the relations correlate well with parameters indicating the
presence of young stellar populations (Schweizer et al. 1990). In Falc\'on-Barroso et al.
(2002, FB02) we used the relation to show that the stellar populations in a sample of inclined
early-type spirals are generally old. 

In Figs. 5, 6 and 7 we show the central \mgb\ and \hb\ indices of our sample as a
function of the central velocity dispersion $\sigma_{\rm cen}$. 
In the figure are
shown the galaxies of this sample, together with the ellipticals and lenticulars of
Paper VI (at r$_e$/8), and a number of literature samples of early-type spirals 
(see caption).
The black line is a best fit to the ellipticals and S0 galaxies in the Coma cluster
of J\o rgensen et al. (1996). The \mgb\ - $\sigma$ relation of elliptical
galaxies and S0's acts as an upper envelope for the Sa galaxies. Although  some
Sa galaxy  centre measurements lie close to the relation, a significant fraction of the galaxies
falls below it. The same effect is seen for the \hb\ - $\sigma$ relation. 
Here the line of ellipticals and S0 galaxies in Coma is not accurately
known, since the H$\beta$ index of a galaxy is crucially dependent on its correction
for emission, and very few emission-line corrected \hb\ indices have been published
in the literature.  One sees, however, a well-defined lower envelope  in the red
points. Using the argumentation of Schweizer et al. (1992),  the line of galaxies in
Coma would correspond to old stellar populations, while deviations would be caused by
younger stars. The fact that our Sa bulges  mostly lie below the \mgb\ - $\sigma$
relation or above the  \hb\ - $\sigma$ relation would indicate that the centres of Sa
bulges generally are significantly younger than  early-type galaxies in the Coma
cluster.
 
This result appears to contradict the tight Mg$_2$ -- $\sigma$ relation for bulges
found by FB02 and also the relation by Jablonka et al.
(1996). It confirms, however, the results of Prugniel et al. (2001), also described
by FB02, who find several early-type spiral galaxies
lying considerably below the Mg$_2$ -- $\sigma$ relation. Notice that there are 
several S0 galaxies that are far away from the relation defined by elliptical
galaxies, in the same location  as the spirals with the lowest \mgb\ values. 
We have converted the central Mg$_2$ values of Jablonka et al. (1996) 
to \mgb\ using the tight relation of
the Vazdekis et al. (1996) models and plotted them as black crosses. The position of
those black crosses is not very different from our bulges, ellipticals and
lenticulars. The galaxies of Bender, Burstein \& Faber (1993) have been selected to
be lenticulars, so it is no surprise that they agree well with FB02. One should note 
that there is a small offset  for the red points, since
their central \mgb\ indices are in general slightly higher than  the value inside
r$_e$/8. Since the velocity dispersion profiles of the ellipticals and S0 galaxies
are generally rising inward they are probably moving slighly along the line.  In
FB02 some more details of this figure are discussed. 

\onecolumn
\begin{figure}
\begin{center} 
  \includegraphics[draft=false,scale=1,trim=0cm 0.cm
    0cm 0cm]{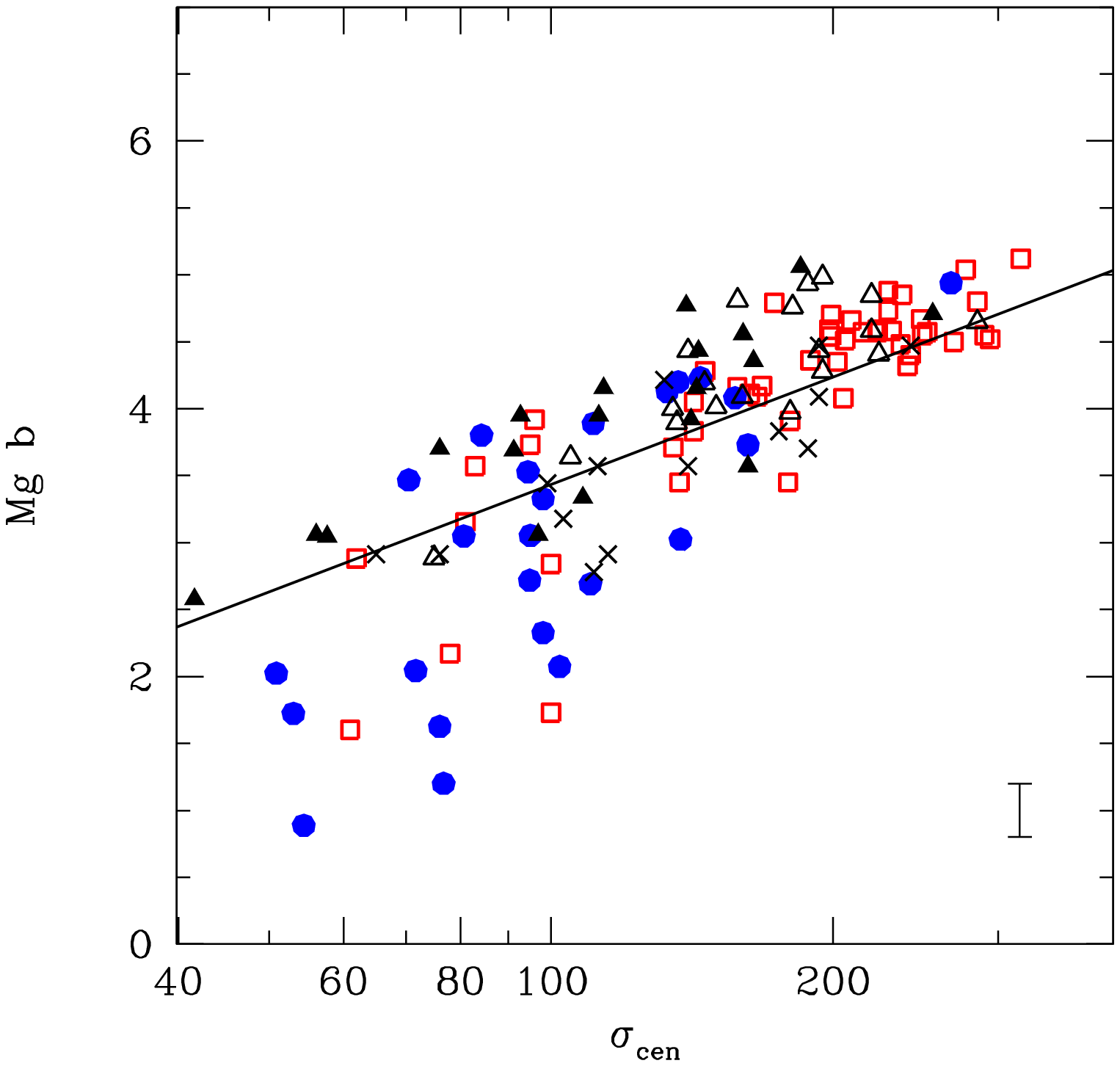} 
\end{center} 
\caption[]{Central \mgb\  as a function of central velocity \
dispersion (in km/s). The open red symbols show the ellipticals and S0 galaxies of Paper VI 
for an aperture of $r_e/8$. The filled dark blue symbols indicate central apertures of the 
galaxies of
this paper  (including a representative error bar). The black line is the least-squares fit to the ellipticals and S0 galaxies in Coma
of J\o rgensen et al. (1996).
As a comparison we also show a few literature samples in black: the filled triangles 
indicate
the highly-inclined bulges of FB02, the open triangles the bulges of
Bender et al. (1993), and the crosses the bulges of Jablonka et al. (1996). For these 3 samples we
have converted Mg$_2$ to \mgb\  using a least-squares fit to all the Vazdekis et al. (1996) models for
which Mg$_2$ $>$ 0.10: \mgb\ = 12.92 Mg$_2$ + 0.537. 
} 
\label{fig:indextype}
\end{figure}

\begin{figure}
\begin{center} 
  \includegraphics[draft=false,scale=1,trim=0cm 0.cm
    0cm 0cm]{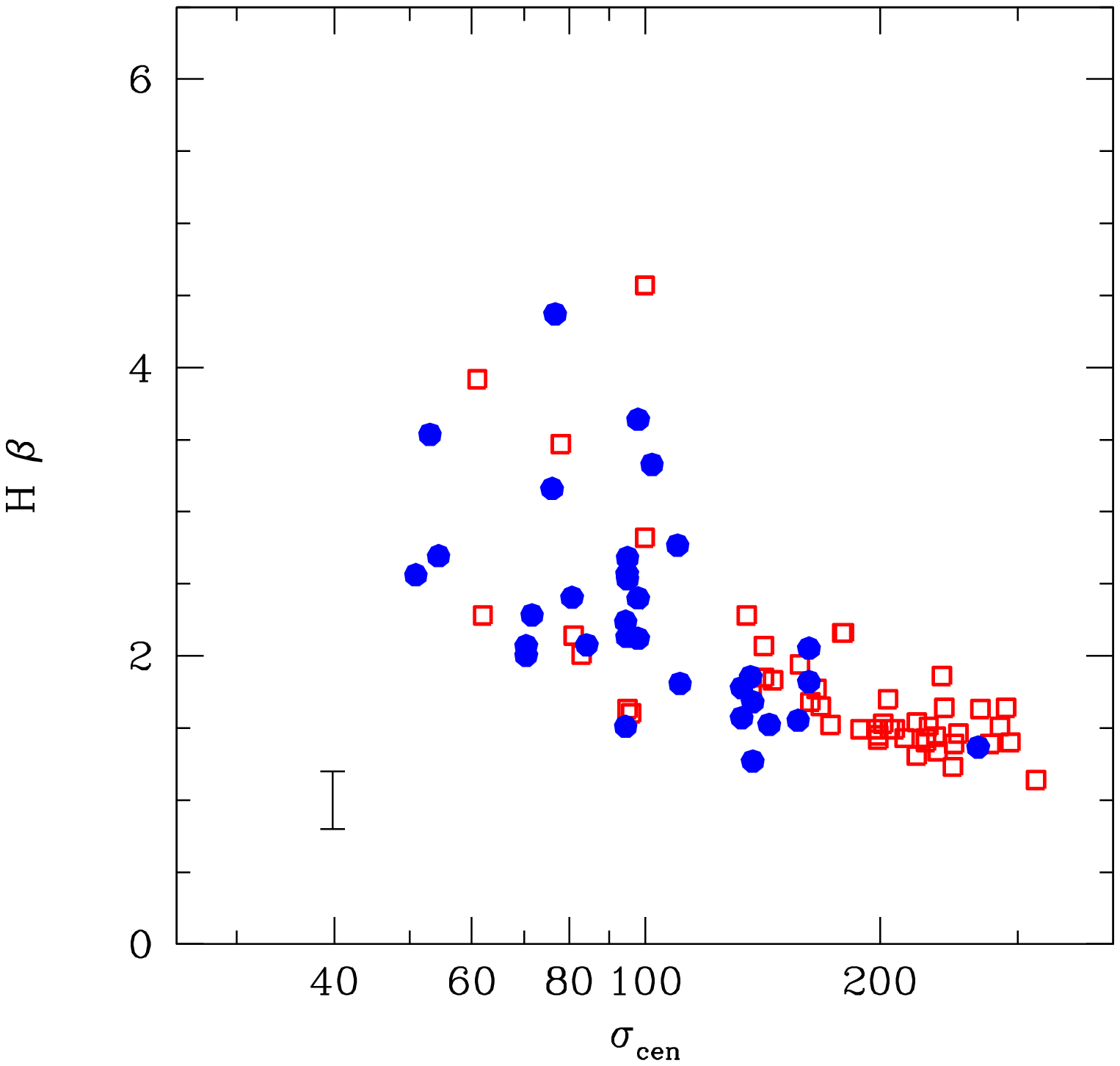} 
\end{center} 
\caption[]{Central H$\beta$ absorption as a function of central velocity \
dispersion $\sigma_{cen}$ (in km/s). The open red symbols show the ellipticals and S0 galaxies of Paper VI 
for an aperture of $r_e/8$. The filled dark blue symbols indicate central apertures of the galaxies of
this paper (including a representative error bar).
} 
\label{fig:indextype}
\end{figure}

\begin{figure}
\caption[]{Same figure as Figures 5 and 6, but now with \hb\ maps of 3 example galaxies, which show
that the stellar populations are young when the galaxies lies off the relation
for galaxies in the Coma cluster.{\bf Figure 7 available as separate jpg-file}}
\label{fig:indextype}
\end{figure}
\twocolumn


In the region of interest ($\sigma$
$<$ 120 km/s) the galaxies of  FB02  generally have higher \mgb\
than the galaxies of this sample. Why this difference? The only important difference
between the two samples is the inclination distribution. If the young stellar populations
would be concentrated in the plane
we would see only old  stellar populations in the inclined sample at 5$\arcsec$ above the 
plane, if the radial extent of the young stellar populations would be limited to the very central regions, 
while this would not be the case for the {\tt SAURON} sample. 
Fortunately, we know more about the sample of FB02. From colours from HST Peletier et al.
(1999) found that the stellar populations at 5$\arcsec$ on the minor axis of all these 
galaxies are old (9 $\pm$ 2 Gyr), except for their 3 Sbc galaxies. 
The fact that we have two-dimensional stellar population information for the {\tt SAURON}
galaxies allows us to understand the position of the points in the \mgb\  - $\sigma$
diagram much better. We have therefore labeled three typical points in this diagram, one 
on the line, and two below it. For these three points we show the \hb\ absorption line 
maps in Figure 7. They indicate that the stellar populations for the galaxy on the line,
NGC 4698, are old, while the \hb\  maps for the two galaxies below the line show
signatures typical of galaxies with young stellar populations. In both the latter galaxies
the regions with young stars are extended. 
Would NGC 4369, if seen at higher inclination, lie on the relation of FB02? Since the 
region dominated by young stellar populations is extended, going out to about 10$''$ on the minor
axis, one would see the old stellar populations at 5$''$ on the minor axis if the inclination
would be larger than 60$^{\rm o}$ and the young populations would be distributed in
a thin disc. NGC 2273 would look old at 5$''$ on the minor axis at any 
inclination angle, since the region of young stars here is small. This qualitative comparison
shows, although not very precisely, that the inclination distribution might be 
the only difference between the sample of FB02 and the {\tt SAURON} Sa sample.
Note that the \mgb\ - $\sigma$ relation
for the Coma cluster is a relation for the oldest galaxies that not necessarily 
all have the same old age of $\sim$ 10 Gyr.
This means that the distance of a galaxy from the line of Coma-galaxies
is a measure of the age-difference between the galaxy and the oldest galaxy at the same
$\sigma$, not of its absolute age.
Since differences in metallicity almost do not affect \hb\ (e.g. Paper VI) the \hb\ -
$\sigma$ diagram is a much cleaner diagram to study these age-differences (Fig.~6). 
Here one can also see the large spread in age for galaxies with 
low central velocity dispersion. A comparison with the sample of FB02 is unfortunately
not possible here.

One might wonder what determines the stellar populations in the centre. For
elliptical galaxies there is a strong relation between the total luminosity and the
central Mg$_2$ index (Guzm\'an et al. 1992, J\o rgensen et al. 1996), or the central
velocity dispersion. Since also $M/L$ correlates with luminosity (e.g. Bender et al.
1993) there is a
strong correlation between the total galaxy mass and the central stellar populations.
It is not clear whether a similar relation also holds for spiral galaxies, although
there is a tight relation between the central  metallicity from the ionised gas and
the mass of spiral galaxies (Zaritsky et al. 1994). For our objects, the \mgb\
-- M$_B$ relation (not shown) is much less tight than the \mgb\ -- $\sigma_{cen}$ relation, 
implying that the
central stellar populations are probably determined more by the local concentration of
matter than by the mass of the galaxy itself. In the same way, there is no good correlation
between \mgb\ and W$_{20}^c$, the inclination-corrected
velocity width of the 21cm profile, generally considered a good mass indicator of galaxies.
Or alternatively, one could say that
central stellar populations are determined more by the central regions than by the galaxy itself.
This indicates that the stellar populations in the bulge are probably determined 
more locally, e.g. by the Star Formation mode, than
globally by the mass of the whole galaxy. This would be a relation similar to the one between
bulge mass and black hole mass (e.g.,
Tremaine et al. 2002) which is tighter than the relation of black hole mass and total galaxy
mass.

\subsection{SSP age as a function of $\sigma$}

Up to this point most of the analysis is based purely on data, without invoking any
stellar population modeling. We now look at
the results of the SSP-analysis. In Fig. 8 we have plotted the central SSP-age
versus the central velocity dispersion. Surprisingly, we find that below $\sigma$ =
100 km/s none of the galaxies is old, and
that the average age drops very fast when going to fainter galaxies. 
Compared to the early-type galaxies of Paper VI Sa galaxies are slightly older
for a given $\sigma_{cen}$ or have a lower central velocity dispersion for a given
age. One finds, however, an opposite effect when comparing the Sa galaxies with the elliptical
galaxies in the Virgo cluster of Yamada et al. (2006), which for the same central velocity 
dispersion are younger.  We might tentatively conclude that the morphological type
of the galaxy is less tightly correlated with the central stellar populations than
the central velocity dispersion, for the types of galaxies concerned (ellipticals, S0s and Sa
galaxies). There are early-type galaxies with young central stellar populations, and also
Sa galaxies. In the same way there are Sa galaxies with old central stellar populations. 
We see that for a given velocity dispersion the Virgo ellipticals are slightly older than
our cluster Sa's, which in turn are slightly older than our field Sa's. The apparent differences 
in age with the sample of Yamada et al.
might also be partly due to the slightly different method used to derive ages.
Interesting also is that ages at 5$\arcsec$ from the centre on the minor axis are
generally older than central ages, strengthening our interpretation of Figs. 5, 6 and 7.
If we use the central velocity dispersion as an indicator of mass, we find that
our trend of galaxy mass with age is similar to the result of
Kauffmann et al. (2003), based on Sloan Digital Sky Survey data, which says that there
are two groups of galaxies, with the less massive galaxies much younger than the
massive ones, separated at  about 3$\times$10$^{10}$ M$_\odot$. 
For the spirals galaxies as a whole we do not have a
clear mass - age relation. Using the inclination-corrected 21-cm velocity width
(Table 1), a
mass-indicator based on the rotation velocity of the gas in the outer disc regions,
we see that a
clear mass-age trend, as indicated by Kauffmann et al. (2003) is probably not present 
(Fig.~9),  although the oldest galaxies are indeed some of the largest ones.
It looks as if bulge mass is a much more important parameter determining
the central stellar populations, rather than total mass. 

\begin{figure}
\begin{center} 
  \includegraphics[draft=false,scale=0.44,trim=0cm 0.cm
    0cm 0cm]{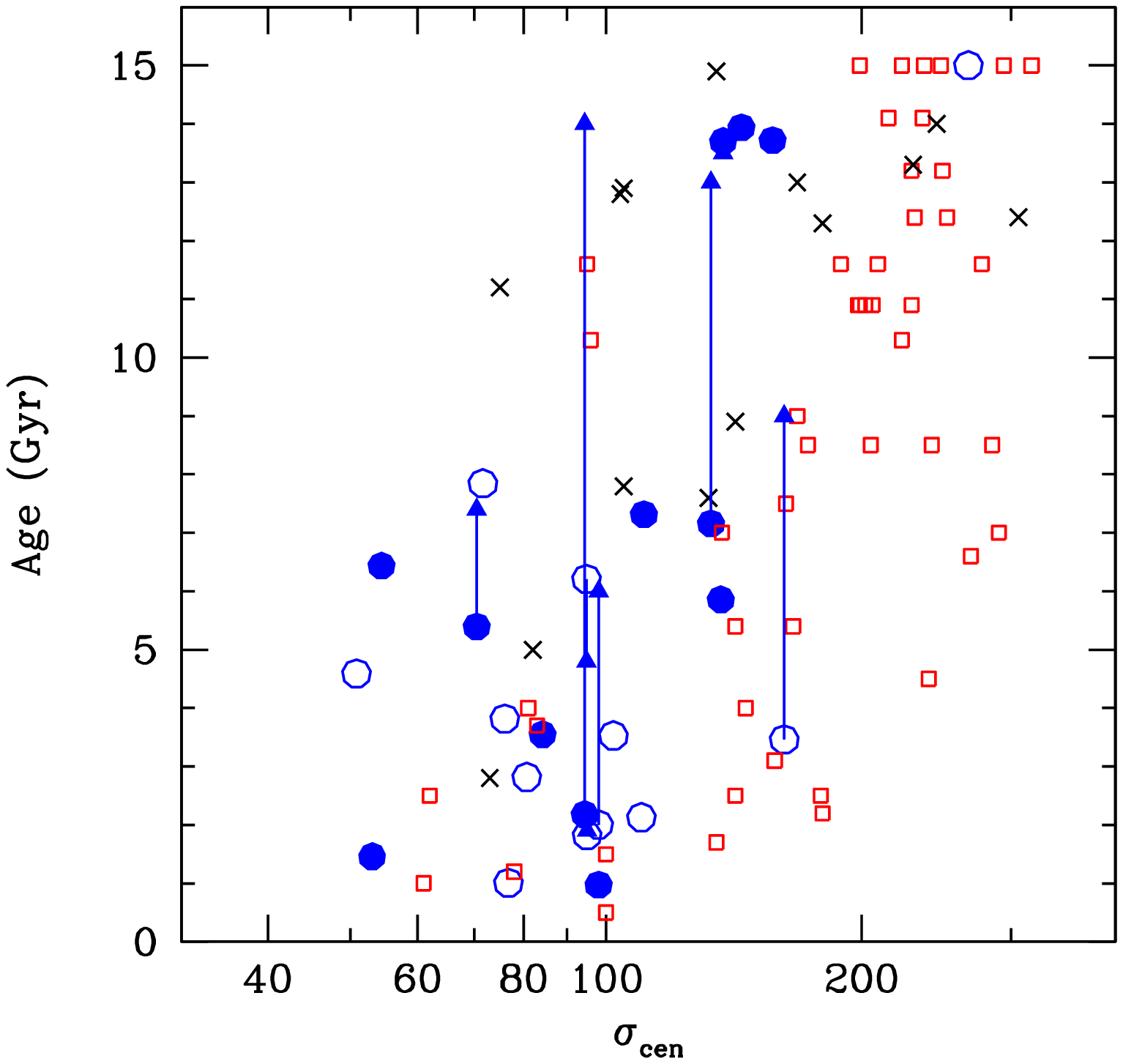} 
\end{center} 
\caption[]{Central SSP-age as a function of central velocity dispersion (in km/s) for
our 12 cluster (blue filled symbols) and field galaxies (open symbols). 
For the galaxies with inclination larger than 60$^{\rm o}$ we have also plotted the
age at 5$\arcsec$ from the centre along the minor axis (arrows). Red open squares are ellipticals and S0s from Paper VI and 
Kuntschner et al. (in preparation). The black crosses are from Yamada et al. (2006), using the ages determined from H$\beta$.
}
\label{fig:indextype}
\end{figure}

\begin{figure}
\begin{center} 
  \includegraphics[draft=false,scale=0.44,trim=0cm 0.cm
    0cm 0cm]{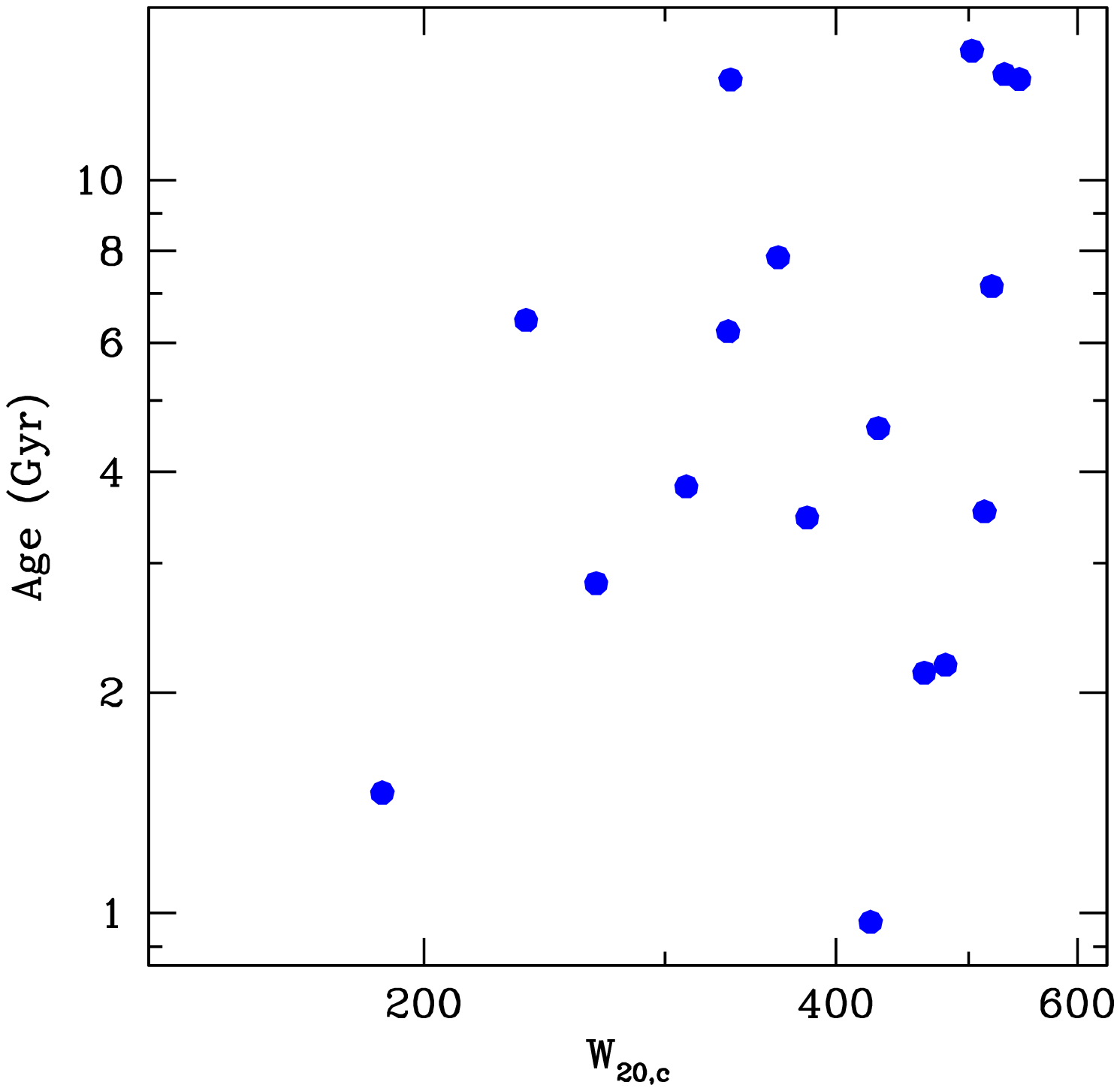} 
\end{center} 
\caption[]{
Central SSP-age as a function of W$_{20}^c$, a the inclination-corrected
velocity width of the 21cm profile (in km/s), a good mass indicator of the galaxy.
Note that W$_{20}^c$ is not available for all galaxies.}
\label{fig:indextype}
\end{figure}

\subsection{Sigma drops}

Central velocity dispersion minima (i.e., sigma drops) are rather common in our
sample (Paper VII).   In this paper we observe distinct dispersion drops in 11
out of 24 galaxies (46\%). This is probably a lower limit considering the medium
spatial resolution of our {\tt SAURON} data. In Table 1 we give the ratio of the 
velocity dispersion in the central aperture of radius 1$\farcs$2 and 
the maximum stellar velocity dispersion in the {\tt SAURON} field. For 14 of our 24 galaxies 
this ratio is significantly lower than 1 (lower than 0.96). For three of those,
NGC 2844, 4245 and 4293, the scatter in the velocity dispersions in the individual 
bins is such that one discovers the sigma drop only after a more careful analysis of the 
velocity dispersion map, and for that reason they were not mentioned in paper VII.
In later type galaxies this fraction is
higher than 46\%. In a paper by Ganda et al. (2006) we show 
that the velocity dispersion
profiles of galaxies of morphological type  Sbc and later are generally rising across
the whole {\tt SAURON} field.  As a comparison, Chung \& Bureau (2004) revealed a
sigma drop frequency of about 40\% in a sample of 30 nearly edge-on S0-Sbc
galaxies, which is consistent with our findings. The first observed cases of central
velocity dispersion minima date back to the late 80s and early 90s (e.g.,
Bottema 1989, 1993). However, this subject has started to gain
significant attention in the last years.  Emsellem et al. (2001) proposed that these were the result of gas forming stars in central discs (see also
M\'arquez et al. 2003, and Emsellem 2006 for a recent review). The first modeling of this effect was presented in Wozniak et al. (2003). Here, the 2-dimensional data clearly confirm that the  $\sigma$ drops are caused by
central discs (Paper VII).  Since sigma drops are rarely seen in
ellipticals, but common in spirals, this is one of the aspects in which spirals and
ellipticals differ. Sigma drops could be caused by stars forming in central gas
discs later on in the evolution of the galaxy. These discs are formed as dynamically cold
systems, and slowly heat up.  As long as they are cold, 
and are responsible for a significant fraction of the light, they will produce
$\sigma$ drops at any inclination, if they are dominating the light.  This
interpretation is consistent with  N-body and SPH simulations of, e.g.,  Wozniak \&
Champavert (2006), who form discs from  gas inflow towards the central regions of
the galaxy and subsequent star formation. Note for example the central regions of NGC~4274. 
In the star formation ring the velocity dispersion is low, since the light is dominated by
a young, cold, disc, causing the central sigma-drop.  In the outer regions, where the
velocity dispersion is higher, the light mainly comes from old stars.

Having determined the stellar population age, we investigate how long a central
disc that causes a $\sigma$-dip can survive.  To quantify the sigma
drops we azimuthally averaged the velocity dispersion along isophotes (in the same
way as was done in Ganda et al. 2006) and used that to  normalize the central
velocity dispersion. We then plot the square root of the quadratic difference between the maximum and 
central velocity dispersion in Fig.~10  (with numbers tabulated in Table 1). This is a measure of the size of the central dip. 
If no dip is present this parameter is equal to zero.
Here we see an interesting effect.  Although the
fraction of $\sigma$-drops for young galaxies is larger, very old central discs exist
as well. The oldest galaxy with a $\sigma$-drop is NGC 4235. This galaxy  has a
Seyfert 1 nucleus, which shows up as a  lower \mgb\ and \hb\ value in the
very centre. So, it is likely that the stellar age calculated there is not very
accurate. However, in the rest of the inner disc (Paper VII) old ages are found
(Figure 1). We conclude that central discs can survive long. This is consistent with
the simulations of Wozniak \& Champavert (2006), who, however, stop their simulation
at 2.1 Gyr, but report that  the amplitude of the $\sigma$-drop had remained the same
for the last Gyr.   Paper VIII shows the age distribution of kinematically-decoupled
cores (KDCs). All large-scale KDCs ($\sim$1 kpc), present in slow-rotating galaxies,
appear old, while most small-scale KDCs ($\le$ 300 pc), present in fast-rotating
galaxies, appear young. Their interpretation is that they are also discs, which like any
stellar population will slowly fade, until the surface brightness is low enough to be
totally overcome by the main galaxy body. The sigma drops we see here are extended
(size $>$ 5$\arcsec$), and mostly older than 1 Gyr. They might be similar to
objects such as the central discs in giant ellipticals (see a discussion in paper VIII).

\begin{figure}
\begin{center} 
  \includegraphics[draft=false,scale=0.44,trim=0cm 0.cm
    0cm 0cm]{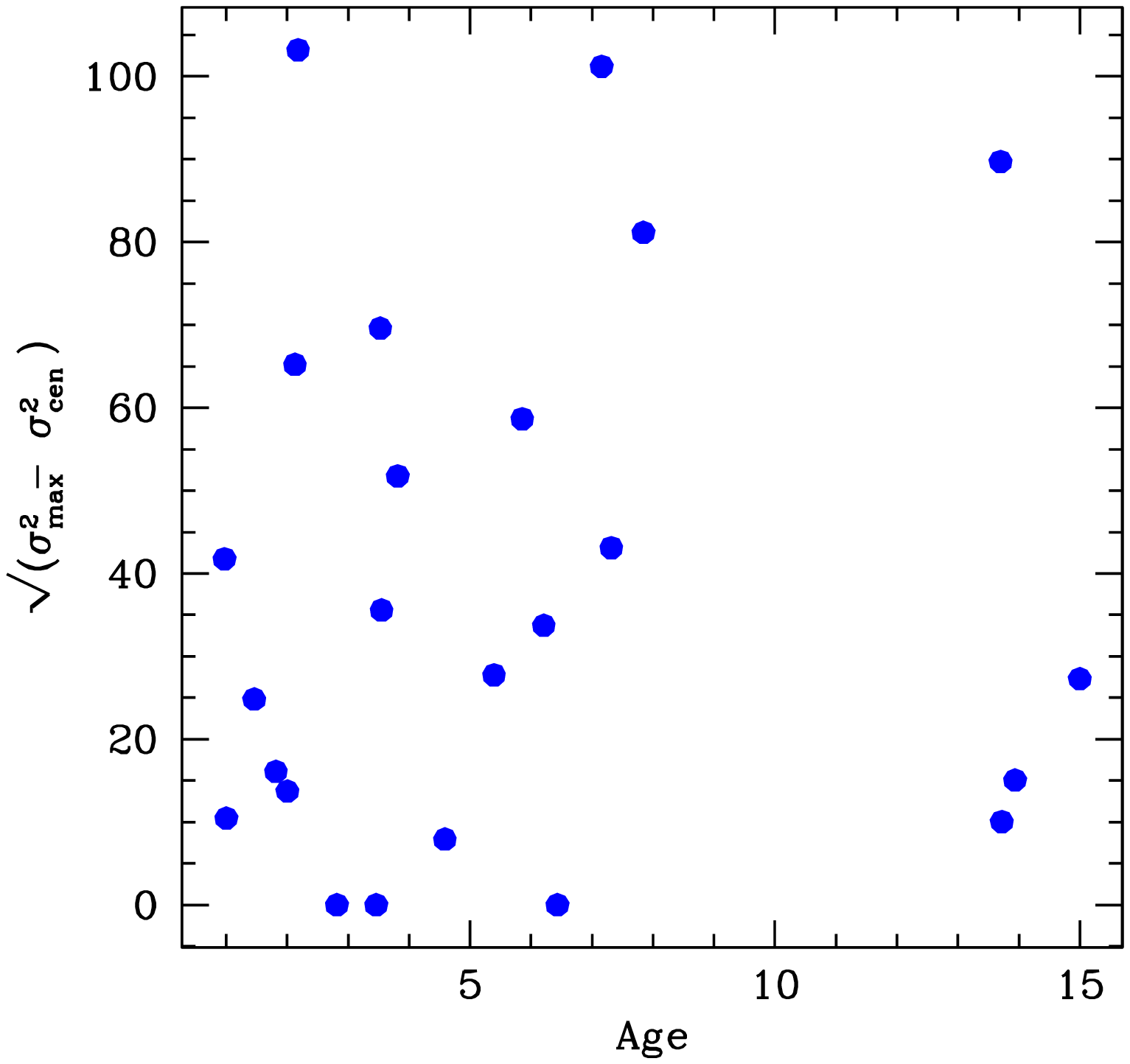} 
\end{center} 
\caption[]{Relation between central SSP age (in Gyr) and the size of the central sigma drop,
indicated by the square root of the quadratic difference between $\sigma_{max}$ and $\sigma_{cen}$, where 
$\sigma_{cen}$ has been calculated in a
central aperature of radius 1.2 arcsec and $\sigma_{max}$ the maximum velocity
dispersion in the azimuthally averaged profile. }
\label{fig:indextype}
\end{figure}

\section{Discussion}

\subsection{Star Formation Modes}

In Section 5 we have found that stellar populations in the centres of early-type spirals seem to
consist of an underlying old stellar population, together with localized younger stellar
populations, either in central discs, rings or more irregular structures. Is this consistent with
what we know from previous work?

Current wisdom in the literature is that there are two modes of star formation (Kennicutt
1998). The first manifests itself by a strong relation between the  morphological type and
the amount of star formation, as measured from H$\alpha$. This causes the average star
formation rate to increase monotonically with morphological T-type. There is however a real
scatter in the star formation per galaxy of about a factor 10. Several factors contribute
to the variations in the Star Formation Rate (SFR), including variations in gas content,
nuclear emission, interactions, and possibly short-term variations in the SFR within
individual objects. 

The second mode can be found in the circumnuclear regions of many spiral galaxies, which harbor
luminous star-forming regions. They have properties that are largely decoupled from those of the more
extended star-forming discs (Morgan 1958, S\'ersic \& Pastoriza 1967). The physical conditions in
the circumnuclear star-forming discs are distinct in many respects from the more extended
star-forming discs of spiral galaxies. The circumnuclear star formation is especially distinctive
in terms of the absolute range in SFRs, the much higher spatial concentrations of gas and stars,
and its burst-like nature (in luminous systems) (Kennicutt 1998).

In contrast to the extended star formation in discs, which varies dramatically along the Hubble
sequence, circumnuclear star formation occurs mostly in early-type spiral galaxies. Stauffer (1982),
Keel (1983), Ho et al. (1997) investigated the dependence of nuclear H$\alpha$ emission in star
forming nuclei as a function of galaxy type. They found that the detection frequency of HII region nuclei is a
strong monotonic function of type, increasing from 0\% in elliptical galaxies to 8\% in S0, 22\%
in Sa, 51\% in Sb, and 80\% in Sc - Im galaxies, although these fractions may be
influenced somewhat by AGN contamination. Among the galaxies with nuclear star formation, the
H$\alpha$ luminosities show the opposite trend; the average extinction-corrected luminosity of
HII region nuclei in S0 - Sbc galaxies is nine times higher than in Sc galaxies.
Thus, the bulk of the total nuclear star formation in galaxies is weighted toward the earlier
Hubble types, even though the frequency of occurence is higher in the late types. Kormendy \&
Kennicutt (2004) report that the central star formation  accounts for 10 - 100\% of the total SFR
of spiral galaxies. The highest fractions occur in early-type spiral  galaxies, which typically
have low SFRs in their outer discs (Kennicutt \& Kent 1983).

Although the picture described above was derived primarily from H$\alpha$ images,  we see exactly
the behaviour described above in our current sample.  Our galaxies, all early-type spirals, and
most of them of type Sa, show a much larger range in age than ellipticals and S0s, derived both
from the index-index diagrams and from the Mg (or \hb\ )- $\sigma$ diagram. Also spatially, one
sees that the younger stellar populations are concentrated near the centre or in annuli
suggestive of resonance rings (e.g. Byrd et al. 1994). The picture is
consistent with all star formation taking place in the disc, and close to the centre.

We find that 29 $\pm$ 9 \% of our galaxies shows younger stars in rings. This number
is compatible with Knapen et al. (2006), who find 30 $\pm$ 5\%. The real fraction might be 
higher, since the detection rate also depends on the spatial resolution.
The fact that galaxies form stars in rings is not new
(see e.g. Benedict et al. 1992 for the beautiful case of NGC 4314 and Knapen et
al. 1995 for NGC 4321). Gas surface densities often peak near the ILR (Inner
Lindblad Resonance, Combes et al. 1992), lending support to the idea that the 
inward flow of gas along the bar slows down and as a result the gas piles up
between the OILR and the IILR (Outer and Inner Inner Lindblad Resonance,
Shlosman et al. 1989).  Inner star forming rings are seen in many,
predominantly barred, galaxies (Knapen et al. 2006). They find that a large
fraction of spiral galaxies show central or circumnuclear H$\alpha$ emission,
indicating the  presence of ionising O or B stars. The galaxies with rings are
predominantly found  in galaxies of types Sa-Sbc, while central star formation
in later type galaxies is generally  patchy.

We also find that the galaxies with lower $\sigma$ values, or lower values of \mgb\, i.e.
generally the smaller galaxies, show a larger range in age than the larger ones. This is
not just the case for the Sa galaxies, but also for the earlier types, mostly S0 galaxies.
One can find evidence for this effect in several places in the literature. 
Yamada et al. (2006), from a study of elliptical galaxies in the Virgo cluster, find that
the scatter in age is much larger at low $\sigma$ than for the most massive galaxies. 
SDSS results for early-type galaxies (Gallazzi et al. 2006) show the same result.
Gavazzi \&
Scodeggio (1996) and Gavazzi et al (1996) compiled UV, visible, and near-IR photometry for
over 900 nearby galaxies. They found an anti-correlation between the SFR per unit mass
and the galaxy luminosity, as indicated by broadband colors and H$\alpha$ EWs. At least
part of this trend seems to reflect the same dependence of SFR on Hubble type discussed
above, but a mass dependence is also observed among galaxies of the same Hubble type.
Also, a similar effect is seen in early-type galaxies. Kuntschner (2000) finds for
elliptical and S0 galaxies in the Fornax cluster that the range in ages for faint galaxies
is much larger than for the large galaxies, which all happen to be old. The fainter
galaxies also seem to have a different star formation mode. NGC 4369, 4383 or 4405, all
fainter than M$_B$=--19, have star formation concentrated in a large central region, and
not in rings, as is much more common is brighter galaxies. For spirals Bell \& de Jong
(2000) find a highly significant correlation between the K-band absolute magnitude and
age, and also between the gas fraction and age, in the sense that fainter galaxies are
younger and have a larger gas fraction (see also Kauffmann et al. 2003). They also find
that there is a kind of 'saturation' in the stellar metallicities: the metallicity of
galaxies with an absolute K-band magnitude of --22 is very similar to the metallicity of
galaxies with an absolute K-band magnitude of --26. The metallicity of galaxies fainter
than an absolute K-band magnitude of --22 can be much lower.

\subsection{Star Formation, Bars and Bulge Formation}

In Section 6.1 we found that stellar populations in the central regions of highly-inclined
galaxies behave differently from those in more face-on galaxies. From that we concluded
that young stars must form in the plane of the galaxy, with a small scale height, rather
than in the whole bulge. For these Sa galaxies, 
classified on the basis of their large bulge, this can only mean that in the radial region
in which the central disc stars dominate, they are responsible for much, or most of the
light that is usually attributed to the bulge  (e.g.\ when doing a bulge-disc
decomposition, e.g. Andredakis et al. 1995, de Jong 1996, McArthur et al. 2003). 

An instructive example is the galaxy NGC 3623, which has a bright {\it thin} 
inner disc
(see e.g. Paper VII), dominating the central stellar populations. The stellar kinematics
show fast rotation and low velocity dispersion in this disc, i.e. disc-like kinematics.
Outside the disc, the rotation drops and the velocity dispersion goes up. Obviously the
central component, when doing the decomposition, is part of the bulge. 
Its appearance and
kinematics however indicate that it is a disc. Our sample contains more cases like this.

Kormendy \& Kennicutt (2004) repeatedly note that the central light distributions of some
galaxies are  very flat, based on observed axial ratios or spiral structure (see also
Kormendy et al. 2006). Figure 8 in Kormendy (1993) shows that a majority of bulges appear
rounder than their associated discs. These include the well-known classical bulges in M31,
M81, NGC 2841,  NGC 3115, and NGC 4594 (the Sombrero galaxy). Some bulges have apparent
flattenings that are similar  to those of their associated discs, as Kent (1985) noted.
Some bulges, however, appear more flattened than  their associated discs; these may be
inner bars. Fathi \& Peletier (2003) show that there is tentative evidence that in
late-type spirals the flattening of bulges is  larger than in early-type spirals.

When one considers the nuclear and circumnuclear star formation, one finds a strong
correlation  with bar structure, and the virtual absence of any other dependence on
morphological type (Kennicutt 1998). This implies that the evolution of the
circumnuclear region is largely decoupled from that of the disc at larger radii. The
strong distinctions between early-type and late-type barred galaxies appear to be
associated with the structural and dynamical properties of the bars. Bars in
bulge-dominated, early-type spirals tend to be very strong and efficient at
transporting gas from the disc into the central regions, while bars in late-type
galaxies are much weaker and are predicted to be much less efficient in transporting
gas (e.g. Athanassoula 1992, Friedli \& Benz 1995). All of the results are consistent
with a general picture in which the circumnuclear SFRs of galaxies are determined
largely by the rate of gas transport into the inner regions. Kormendy \& Kennicutt
(2004) claim that in this way so called pseudo-bulges are formed, components that
dominate the light in the centre, but have disc-like properties. They estimate a
typical formation timescale for the central discs of 1 Gyr.  Since circumnuclear
star-forming rings of this type are seen in 10\% of intermediate-type spiral galaxies
(S\'ersic 1973, Maoz et al. 1996) they suggest that about half of unbarred spirals
and nearly all barred spirals may have formed central disks in this mass range.

According to the current literature there are several kinds of bulges.
Bulges that photometrically (r$^{1/4}$ surface brightness law) and kinematically (slowly
rotating, but with high $\lambda_R$, Emsellem et al. 2007) resemble elliptical galaxies are often called classical
bulges (Kormendy \& Kennicutt 2004). A bulge
consisting only of the fast-rotating component is called a pseudo-bulge in this reference.
In making this classification Kormendy \& Kennicutt rather simplify things, since e.g.
Balcells et al. (2003) show that very few bulges have real r$^{1/4}$ surface brightness 
profiles.  Athanassoula (2005) claims that there are three types of bulges: the classical 
bulges, which form by collapse or merging, 
disc-like bulges, which result from the inflow of (mainly) gas to the  centre-most parts,
and subsequent star formation, and
boxy and peanut bulges, which are seen in 
near-to-edge-on galaxies and which are in fact just a part of the bar seen edge-on,  and
therefore not part of the bulge in the definition of this paper (Section 1).
Here we add another piece of the puzzle. From the stellar population distribution, 
by comparing a sample uniformly distributed in inclination with a sample biased towards 
high inclination we infer that galactic bulges have more than
one physical component: generally they have a slowly-rotating, elliptical-like component,
and one or more fast-rotating components in the plane of the galaxy.  This picture also
nicely explains the fact that bulge populations in general are very similar to those in the
disc (e.g. Peletier \& Balcells 1996, Terndrup et al. 1994).  
Fisher (2006) finds one more
piece of evidence. In galaxies which he classifies as having a disc-like bulge the central
star formation rate, as obtained from the Spitzer 3.6 - 8.0 $\mu$m colour, is higher than
in galaxies with classical bulges. This shows that the Spitzer colour measures similar
things as our optical line indices. Our conclusions here is that we see classical and
disc-like bulges, which in many cases co-exist in the same galaxy. 

The Galactic Bulge also fits well into this picture. Zoccali et al. (2003) find that stars
in the Galactic Bulge, measured 6 degrees above the Galactic plane, are as old as Galactic
globular clusters, at least 10 Gyr. This indicates that our Galaxy probably has an old
central bulge component, maybe similar to a classical bulge. We do not know of a component
with disc-like kinematics in our Galaxy, but near the Galactic centre stars are currently
being formed at a fast rate.  The Galaxy has a significant bar, though (Binney et al.
1991; Weiland et al. 1994).  Also, the Galaxy has a prominent molecular ring, and the
radial velocity dispersion of the  bulge and inner disc are the same (~100 km/s; Lewis \&
Freeman 1989; Spaenhauer, Jones \& Whitford 1992).  

In the centres of almost all nearby
early-type spirals, evidence is also seen for the presence of an inner disc-like component
(Peletier et al. 1999). Dust extinction is detected in 95\% of their sample, sometimes
associated with blue light, indicating star formation. It is possible that many galaxies
contain small star forming discs with diameters of about 100 pc in their centre, on top of
central star clusters (e.g. B\"oker et al. 2002, Carollo et al. 1998, Balcells et al.
2003). These components are also disc-like components, photometrically belonging to the bulge.
Bulges often have several dynamical
components (see e.g. Paper VII, Falc\'on-Barroso et al. 2003), with different kinematics,
so it looks as if bulges are not just disc-like or elliptical-like,
but consist of several components. The current sample indicates that,  at least in Sa
galaxies, both disc-like and elliptical-like or classical bulges often coexist.

Coming back to the definition of bulges: if one would define a bulge as a kinematically slowly
rotating system, one would find that in the central regions of many Sa
galaxies the light would be dominated by another component, with disk-like properties, being
flatter, and rotating faster. If one would adopt a morphology-based definition, and take as the
bulge the component that is boxy or peanut-shaped, one would get a component that also includes the
thick part of an inner bar, i.e. a fast rotating disk component (Kuijken \& Merrifield 1995, Chung
\& Bureau 2004, Bureau \& Athanassoula 2005). It is therefore important to understand what
definition is used when discussing bulge properties.

\section{Conclusions}

The main conclusions of this paper are:

We have presented absorption line strength maps in \hb, \fe, and \mgb\ of a sample of $24$
representative early-type spiral galaxies, mostly of type Sa.
This paper constitutes the first
large sample of spirals with spatially  resolved absorption line strengths.

\begin{enumerate}
\item The absorption line maps show that many galaxies contain some younger
populations, formed in mini-starbursts in small or large inner discs
(100-1000 pc), or in circumnuclear star forming rings, often related to bars.
These mini-starburst cause a considerable scatter in index-index diagrams such
as \mgb\ -- \hb\ and \mgb\ -- \fe\ , more than is measured for early-type
galaxies. As a result, there is not only a wide range in ages, even within the
galaxies, but also in abundance ratio Mg/Fe. The different star formation
modes (starbursting and more quiescent) are also reflected in the large 
range of Mg/Fe that is observed.

\item All of our galaxies lie on or below the \mgb\ -- $\sigma$ relation for
elliptical galaxies in the Coma cluster by  J\o rgensen et al. (1996). If that
relation is considered to be a relation for old galaxies we see that our sample
of spirals has a considerable scatter in age, with the highest scatter at the
lowest $\sigma$, i.e. for the faintest galaxies.

\item To explain this result, and also the result from the literature that
stellar populations on the dustfree  minor axis of inclined spirals are
generally old, we conclude that star formation in the inner regions 
only occurs in a thin disc. Since this inner disc dominates the light in
many cases, it dominates the light of the bulge.  Above the
plane, the light is dominated by another galaxy component, resembling an elliptical galaxy,
consisting of old stars. In this way one can explain in a natural way the 
different types of bulges given by Kormendy \& Kennicutt (2004) and Athanassoula (2005).

\item Sigma drops are found in about half our sample and are
probably caused by these fast rotating flat central discs. 
The stellar populations in these inner components are
not significantly younger than those in galaxies without them, indicating that
the discs causing the velocity dispersion drops can be long-lived.  

\end{enumerate}

\section*{Acknowledgements}
It is a pleasure to thank the ING staff, in particular Rene Rutten, Tom Gregory
and Chris Benn, for enthusiastic and competent support on La Palma. 
We thank Martin Bureau for useful comments on the manuscript and Kambiz Fathi,
Philippe Prugniel and Marc Balcells for help during the process of this work. 
Alexandre Vazdekis is acknowledged for providing us his new MILES-based models 
in advance of publication, and Rob Proctor for giving us the data of his
papers in electronic format. The
{\tt SAURON} project is made possible through grants 614.13.003 and 781.74.203 from
ASTRON/NWO and financial contributions from the Institut National des Sciences
de l'Univers, the Universit\'e Claude Bernard Lyon~I, the universities of
Durham, Leiden, Groningen and Oxford, the British Council, PPARC grant
`Extragalactic Astronomy \& Cosmology at Durham 1998--2002', and the
Netherlands Research School for Astronomy NOVA.  RLD the award of a Research
Fellowship from the Leverhulme Trust. 

%
%
%

\bibliographystyle{mn2e} 

\clearpage
\appendix
\section{Description of individual galaxies}
\label{sec:galaxies_notes}
Here, we briefly comment on the structures observed and provide some relevant
references in the literature for the galaxies for which we have been
able to find relevant information. These comments are additions to
the contents of Appendix A of Paper VII, where we focused on the stellar and gas
kinematics, as well as the emission line strengths.We only discuss earlier work 
in which the galaxies are treated as resolved
objects.

\begin{description}

\item[\bf NGC 1056] is a rather faint galaxy,  classified in NED as Sy 2.
It is part of the list of Markarian et al. (1989). Central line strength
measurements are also available from Prugniel et al. (2001), who have similar
conclusions from Mg$_2$ and $\langle {\rm Fe} \rangle$. The young central ages are confirmed by low
[OIII]/\hb\ values. Surface photometry in V and I is given by Heraudeau et al.
(1996).

\item[\bf NGC 2273] is a double-barred active (Sy 2) galaxy. Its emission line
distribution has been studied by Ferruit et al. (2000). In the inner regions
the galaxy has a very dusty zone occupied by young stellar populations in a
central disc.  

\item[\bf NGC 2844] is an \hii\ galaxy (NED) showing ongoing star formation,
accompanied by dust extinction, in a rather highly inclined ring. The galaxy
is also part of the sample of Prugniel et al. (2001).

\item[\bf NGC 3623] is a large member (M 65) of the Leo triplet. Its stellar
populations are generally old, with its central disc, clearly seen in the
stellar kinematics. Afanasiev \& Sil'chenko (2005) used our data
for this galaxy, together with line strength maps from the MPFS instrument.
They observe a central drop in \mgb, and also central drops in Fe 5270 and
\hb\ . We do not reproduce their result, and find relatively small radial
gradients. Also, decreases in
\mgb\ are usually coupled with increases in \hb. It is possible that
Afanasiev \& Silchenko applied a much lower emission line correction in the
middle (see  Fig. 1). Line indices are also published by Proctor et al. (2002
and 2000). For a comparison see Section 4. An integrated spectrum is given in
Kennicutt (1992).

\item[\bf NGC 4220] displays a prominent dust lane at a distance of 5\arcsec\
north-east of the centre, with younger stellar populations in the dust lane.

\item[\bf NGC 4235] is a nearly edge-on  Seyfert 1.2 galaxy. In the line strength
maps one can see the continuum emission of the central AGN in the very
centre. Jim\'enez-Benito et al. (2000) present Fe, Mg and Ca absorption line
strengths and find that the Ca triplet lines are stronger than expected from the
other lines. This is likely to be due to the non-stellar continuum.

\item[\bf NGC 4245] is a barred galaxy with a prominent dust and star formation
ring in the centre (Erwin \& Sparke 2003). Its central STIS spectrum was
analysed by Sarzi et al. (2005b), who found that the galaxy has an old stellar
population together with a small fraction of stars of age 10$^{8.5}$ years old.
M\"ollenhoff \& Heidt (2001) perform B/D decompositions for this galaxy in many
passbands. 

\item[\bf NGC 4274] is a double barred galaxy (Erwin 2004) with
significant amounts of dust in the inner parts. The kinematics here 
is dominated by an inner disc, which is also prominent in molecular CO
(Koda et al. 2005). Young stars are formed mainly in a ring around this disc.

\item[\bf NGC 4293] is a relatively highly-inclined galaxy with a strong dust lane 
passing close to the nucleus and another dust lane about 7\arcsec\ south of 
the centre. Young stars are seen in the dust lane closest to the nucleus. As for
several of the galaxies of this sample, driftscan spectra are presented by
Gavazzi et al. (2003).

\item[\bf NGC 4314] is a low-inclined, well studied barred galaxy in the Virgo
cluster (Benedict et al. 1996,  P\'erez-Ram\'\i rez et al. 2000), mainly known
for a inner star-forming ring in the inner 10\arcsec. Benedict et al.
analyse ground-based (1992) and HST-images  (1993) of this object. They detect
a inner star forming ring in optical and near-IR  colors, together with a
inner spiral just exterior of the ring. They conclude from   the optical and
near-infrared photometry that a pattern of younger stars exist, stronger  at the
ring, and almost disappearing at the linear dust lanes. Other
surface photometry was presented by Friedli et al. (1996) and Wozniak et
al. (1995). Sarzi et al. (2005b) analyse blue HST STIS-spectra of the centre of
this galaxy. They find that the stellar populations are consistent with an old
age (10$^{10}$ years). Gonz\'alez-Delgado et al. (2004) measure absorption line
indices from central  STIS spectra, and classify it as a galaxy with old
stellar populations and weak [OI]. The galaxy looks very similar to NGC 4321
(M100), for which a detailed stellar population analysis based on {\tt SAURON} data
is presented by Allard et al. (2006).

\item[\bf NGC 4369] is a faint low-inclination galaxy. The central regions are
strongly affected by dust (Falc\'on-Barroso et al. 2006). Ho et al. (1997) classify the 
nuclear spectrum as an HII nucleus, indicating a circumnuclear region with star
formation activity but without AGN. Koda et al. (2005) show that a significant amount of
molecular gas (about 10$^8$ M$_\odot$) is found in the central regions of this galaxy.

\item[\bf NGC 4383] is another starburst galaxy in the Virgo cluster (Rubin et al. 1999).
Rubin et al. state that this galaxy has an unusually shallow nuclear velocity rise, confirmed
in Paper VII. 
The H$\alpha$ + [N~II] image of NGC 4383 shows filaments of ionised gas, suggesting 
significant noncircular gas motions due to the starburst (Koopmann et al. 2004).
The {\tt SAURON} ionised gas velocity field is also possibly suggestive of a starburst outflow. 
The [OIII] emission in this galaxy is so large that we have some difficulty removing 
all of it, resulting in very low Fe 5015 absorption line indices. Despite of this 
the light in the inner regions is dominated by young stellar populations.
\looseness-2

\item[\bf NGC 4405] is another faint galaxy with an HII nucleus. This
galaxy also has a considerable amount of extinction in the central regions, although
less than NGC 4369 or NGC 4383.
The \hbeta\ gas map displays several {\it clumps} around the centre indicating the 
starbursting nature of the star formation in this galaxy. The line indices show 
generally young stellar populations in the central regions. \looseness-2

\item[\bf NGC 4425] is a highly inclined galaxy without much dust with a boxy bulge component 
(Paper VII). The ionised-gas maps show only
patchy traces of emission. The absorption line maps show that the stellar 
populations are mainly old. \looseness-2

\item[\bf NGC 4596] is a non-interacting, strongly barred, galaxy (Gerssen et al. 1999).
The {\tt SAURON} stellar velocity field displays regular rotation along an axis
misaligned with respect to the photometric major axis, due to the
presence of a strong bar. The galaxy has an inner disc which dominates the inner 
5\arcsec\ (Paper VII). The galaxy has little ionised gas, and its stellar populations
from the {\tt SAURON} absorption lines are predominantly old. Gonz\'alez-Delgado et al. (2004), from
central STIS absorption line spectra, classify this galaxy as a galaxy with strong [OI] lines and old
stellar populations. Sarzi et al. (2005b), from the same spectra, confirm this result.
\looseness-2
 
\item[\bf NGC 4698] might be considered the example of the early-to-intermediate Sa
galaxy (Sandage \& Bedke 1994). It contains a  low-luminosity Seyfert 2 nucleus 
(Ho et al. 1997). The stellar kinematics is rather unusual (Paper VII), for example 
because of the presence of a kinematically decoupled central disc, in which the stars rotate 
perpendicularly to the stars elsewhere in the central regions (
Pizzella et al. 2002, Sarzi et al. 2000). The galaxy has little ionised gas, and its stellar populations
from the {\tt SAURON} absorption lines are predominantly old. A stellar population analysis
from STIS spectra also shows that the central stellar populations are old.
\looseness-2

\item[\bf NGC 4772] is a an intermediate size galaxy showing regular rotation 
along the major axis. The unsharp-masked
image reveals a small dusty disc in the centre of the galaxy, the orientation of
which seems consistent with that of the ionised gas in the central 5\arcsec, but rotating
almost in the opposite sense of the stars, in agreement with Haynes et al. (2000).
The absorption lines show that the stellar populations are predominantly old
in the {\tt SAURON} field.

\item[\bf NGC 5448] is an active barred galaxy  (L2; Ho et al. 1997) with a dusty central
region. The
stellar kinematics display a regular disc, and signatures of an inner fast
rotating component in the central $\approx$5\arcsec. The ionised-gas kinematics features an
'S-shaped' zero-velocity curve, suggestive of gas radial motions. 
We refer the reader to Fathi et al. (2005)
for a detailed study of this galaxy using {\tt SAURON}. In the absorption line maps one 
sees an increase in H$\beta$ and a decrease in \mgb\ in the central 5\arcsec. This 
is exactly the region of the central disc, and the excess of dust extinction. \looseness-2

\item[\bf NGC 5475] is an isolated, nearly edge-on galaxy (Balcells \& Peletier 1994). Colour profiles
in optical and near-infrared have been presented in Peletier \& Balcells (1996, 1997). 
The \oiii\ emission extends out mainly along the galaxy minor axis, possibly in a 
polar ring. \looseness-2

\item[\bf NGC 5636] is a faint, strongly barred galaxy in a non-interacting pair with
NGC\,5638 at 2\arcmin, with a  stellar
velocity dispersion that is  one of the lowest in the sample. Our line strength 
maps are rather noisy, indicating overall old stellar populations with younger 
stars near the centre. This agrees with the ionised gas maps, which show very low 
\oiii/\hbeta\ values, suggesting ongoing star formation there. \looseness-2

\item[\bf NGC 5689] is an almost edge-on barred galaxy with a box-shaped bulge
(L\"utticke et al. 2000). A strong dust lane is found at $\approx$5\arcsec\ South 
of the nucleus. Colour maps of Peletier \& Balcells (1997) show that this dust lane 
is also obscuring the centre of the galaxy. Carollo et al. (1997) present an optical
HST-image of this galaxy, although no bulge-disc decomposition is done 
because of the large amount of dust obscuration. Star formation associated to this
central dust lane could be responsible for the lower ages we see from the
absorption line maps. Line strength values of this galaxy were presented by
Proctor et al. (2000).
\looseness-2

\item[\bf NGC 5953] is a  Liner Seyfert 2 galaxy (NED) closely interacting with NGC\,5954
(Reshetnikov 1993, Gonz\'alez-Delgado \& P\'erez 1996). This interactions manifests
itself in a kinematically decoupled stellar core, and ionised gas following the motions
of this KDC, but rotating perpendicular to the stars in the outer regions of the 
{\tt SAURON} field. The galaxy shows a ring of low \oiii/\hbeta\ (Paper VII, Yoshida et 
al. 1993). The stellar absorption
line maps confirm that this is a star formation ring, with low \mgb\ and \fe\ values
and high \hb.  A more detailed discussion of this galaxy is given in Falc\'on-Barroso 
et al. (2006b). \looseness-2

\item[\bf NGC 6501] is part of a group together with NGC\,6467, 
NGC\,6495, PGC\,61102 and NGC\,6500 (Giuricin et al. 2000). It is a massive galaxy,
the most distant of our sample, showing regular rotation and no detection of ionised gas
(Paper VII). The absorption line indices are consistent with old stellar populations,
confirming the results of Cid-Fernandes et al. (2004, 2005), which are based on 
an absorption line study using a larger wavelength range.  \looseness-2

\item[\bf NGC 7742] is a well-known face-on galaxy  (classification T2/L2, Ho et al. 97) hosting a prominent 
star-forming ring surrounding a bright nucleus (see Paper II), and with the 
ionised gas counter-rotating w.r.t. the stars (see also Paper VII). The absorption 
line strength maps show a strong star forming ring, also coindicing with a low \oiii/\hbeta\ emission
line ratio. In the centre, however, we find that the stellar populations are fairly old,
something confirmed by the line strength analysis of Cid-Fernandes et al. (2004, 2005).\looseness-2

\end{description}

\label{lastpage}
\end{document}